\documentclass{article}

\usepackage{graphicx}

\author{Aron C. Wall\footnote{aroncwall@gmail.com}
\\ \textit{Department of Physics} \\ \textit{University of California, Santa Barbara}
\\ \textit{Santa Barbara, CA 93106, USA} }
\title{A proof of the generalized second law for rapidly changing fields and arbitrary horizon slices}
\date{\today}

\begin{document}

\maketitle

\begin{abstract}
The generalized second law is proven for semiclassical quantum fields falling across a causal horizon, minimally coupled to general relativity.  The proof is much more general than previous proofs in that it permits the quantum fields to be rapidly changing with time, and shows that entropy increases when comparing any slice of the horizon to any earlier slice.  The proof requires the existence of an algebra of observables restricted to the horizon, satisfying certain axioms (Determinism, Ultralocality, Local Lorentz Invariance, and Stability).  These axioms are explicitly verified in the case of free fields of various spins, as well as 1+1 conformal field theories.  The validity of the axioms for other interacting theories is discussed.
\newline\newline
PACS numbers: 04.70.Dy, 04.62.+v
\end{abstract}

\newpage
\tableofcontents
\newpage

\section{Introduction}

This article will describe a set of physical assumptions which are sufficient for a semiclassical gravitational theory to obey the generalized second law (GSL) of thermodynamics \cite{hawking75}.  From these physical assumptions, a proof of the GSL will be given for rapidly evolving matter fields and arbitrary horizon slices.  This shows that the GSL holds in differential form, i.e. the entropy is increasing at each spacetime point on the horizon.  As far as I am aware, this is the first time such a general proof of the GSL has been given.

The GSL appears to hold on any causal horizon, i.e. the boundary of the past of any future infinite worldline \cite{JP03}.  Causal horizons include black hole event horizons, as well as Rindler and de Sitter horizons.  The GSL states that on any horizon, the total entropy of fields outside the horizon, plus the total entropy of the horizon itself, must increase as time passes.  This total increasing quantity is known as the generalized entropy.

More precisely, for any complete spatial slice $\Sigma$ intersecting the horizon $H$, the generalized entropy of $\Sigma$ is given by
\begin{equation}
S_\mathrm{H} + S_\mathrm{out}.
\end{equation}
In general relativity, the horizon entropy is proportional to the area\footnote{Because $S_\mathrm{out}$ is a c-number, for consistency it is necessary to interpret $S_\mathrm{H}$ as a c-number as well.  In this article, this will be done by taking the semclassical approximation, in which the area $A$ is a classical quantity, sourced by the expectation value of a quantum operator.  However, this semiclassical approximation can only be an approximation to the true quantum gravity theory, in which the area $A$ becomes an operator.  In Ref. \cite{10proofs} I argued that one should then interpret $A$ as being the expectation value of the quantum area operator.}):
\begin{equation}
S_\mathrm{H} = \frac{A}{4\hbar G}|_{\Sigma\,\cap\,H}.
\end{equation}
The second term is the von Neumann entropy of the matter fields restricted to the region outside of the horizon:
\begin{equation}
S_\mathrm{out} = -\mathrm{tr}(\rho\,\ln\,\rho)|_{\Sigma\,\cap\,I^{-}(H)}.
\end{equation}
However, this outside entropy term has an ultraviolet divergence at the horizon due to the entanglement entropy of fields at very short distances.   So to define the generalized entropy, some kind of renormalization scheme must be employed to subtract off these divergences (cf. section \ref{ren}).

Historically, the laws of thermodynamics for matter have provided substantial clues about the microscopic statistical mechanics of atomic systems.  It seems probable that the GSL will provide similar insight into the statistical mechanics of spacetime itself \cite{sorkin83}.  Because quantum gravity is currently outside of our experimental range of detection, any help which can be obtained from the GSL would be very useful.  The GSL is especially evocative because of how surprising it is: it essentially says that an apparently open system (the exterior of the horizon) behaves in roughly the way that we would expect a closed thermodynamic system to behave.

There are several different claims that in order for the GSL to be true, certain restrictions must hold even semiclassically on e.g. bounds on the entropy and/or number of particle species proposed by Bekenstein \cite{bek81b}, Bousso \cite{bousso02b}, or Dvali \cite{DS08}, bounds on the fine structure constant \cite{davies08}, the unbrokenness of the Lorentz group \cite{EFJW07}, and/or energy conditions \cite{anec}.  If true, these claims hint at important restrictions on any good theory of quantum gravity.  (However, in my opinion, only the last two of these claims have been clearly established.)  One way to test these proposed requirements is by proving the GSL, and thus seeing explicitly what assumptions are necessary.  Once we know what key assumptions are necessary for the GSL to hold semiclassically, we will be in a better position to guess background-free constructions of quantum gravity based on thermodynamic principles.

Until recently, there were satisfactory proofs of the semiclassical GSL only in the `quasi-steady' case in which the fields falling into the black hole are slowly changing with time \cite{10proofs}.  One such quasi-steady argument was the illuminating but incomplete proof by Sorkin \cite{sorkin98} (reviewed in Ref. \cite{10proofs}).  Sorkin considered the case of a physical process $\mathcal{P}$ (which may involve information loss), with the property that a thermal state
\begin{equation}
\sigma = \frac{e^{-\beta H}}{Z}
\end{equation}
evolves to itself under the process:
\begin{equation}
\mathcal{P}(\sigma) = \sigma.
\end{equation}
He then invoked a theorem saying that whenever this happens, the free energy of any other state $\rho$ cannot increase under the same time evolution:
\begin{equation}
(\langle H \rangle - TS)_\rho \ge (\langle H \rangle - TS)_{\mathcal{P}(\rho)}
\end{equation}
The free energy can then be related to the generalized entropy using the ``first law'' of horizon thermodynamics
\begin{equation}
dE = T\,dS_{\mathrm{H}}
\end{equation}
(which applies only to slowly changing horizons).  Unfortunately, the proof founders when applied to black holes \cite{10proofs}, since the state outside the black hole could only be shown to be thermal outside of the bifurcation surface, while a nontrivial application of the GSL requires time evolution from one slice of the horizon to another slice.  Furthermore the Hartle-Hawking thermal state exists only for nonrotating black holes, so there are even worse problems in applying the proof to Kerr black holes.

My previous proof in Ref. \cite{rindler} side-stepped these problems for the special case of (perturbed) Rindler wedges evolving to other Rindler wedges.  In this case it was possible to show that the GSL holds semiclassically even for rapid changes to the horizon, at every instant of time, using a reasonable assumption about the renormalization properties of $S_\mathrm{out}$.  However, this proof was limited to Rindler horizons sliced by flat planes; it was unable to reach de Sitter space, black holes, or even arbitrary slices of Rindler horizons.  The basic problem is that the proof requires not only a boost symmetry of each wedge (in order to show that the state restricted to the wedge is thermal), it also needs a null translation symmetry (so that there will be multiple thermal wedges).  But this is more symmetry than is possessed by most spacetimes with stationary horizons.

In this article I will generalize the proof to (semiclassical perturbations of) arbitrary slices $\Sigma$ of the future horizon $H$.  The new ingredient is the technique of restricting the quantum fields to a null hypersurface.  In particular (at least for free fields) there is an infinite dimensional symmetry group due to the freedom to reparameterize each horizon generator separately \cite{schroer09}.\footnote{This group is isomorphic to the subgroup of the Bondi-Metzner-Sachs group which preserves horizon generators.}  This symmetry will play an important role in the proof of the GSL in section \ref{proofon}.

Restriction to a null surface is helpful for solving a variety of quantum field theory problems, e.g. deep inelastic scattering in QCD, because of the insight it gives into the quantum vacuum \cite{burkardt96}.  The technique was used by Sewell to derive the Hawking effect in a very illuminating way \cite{sewell82}.  More recently, it has also been used as a simple way to characterize quantum fields on Schwarzschild past horizons \cite{DMP09} and future horizons \cite{MP03}, certain past cosmological horizons \cite{DMP08}, 1+1 Rindler horizons \cite{MP04}, de Sitter horizons \cite{pinamonti05} and the conformal boundary of asymptotically flat spacetimes \cite{moretti05}.\footnote{Some of this work refers to this principle of restricting to a null surface by the name of ``holography'', because the null surface has one less dimension than the rest of the spacetime.  But this use of the term is somewhat misleading when compared with the normal usage in quantum gravity, in which it refers to the ability to determine spacetime data from a codimension 2 surface.  Holography in this latter sense should normally only arise when gravitational effects are taken into account.}

The algebra of observables $\mathcal{A}(H)$ on the horizon plays an important role in the proof: it is required to exist and satisfy four axioms described in section \ref{sym}.  In the case of free fields and 1+1 conformal field theories, it will be shown that there exists a horizon algebra satisfying these axioms.

In the case of general interacting quantum field theories, the restriction of the fields to a null hypersurface is a more delicate matter.  Nevertheless, there are reasons to believe that interacting field theories also satisfy the axioms.  At least at the level of formal perturbation theory, the horizon algebra is completely unaffected by the addition of certain kinds of interactions, including both nonderivative couplings, and nonabelian Yang-Mills interactions.  However, renormalization effects can lead to the introduction of additional higher derivative couplings, as well as infinite field strength renormalization.  Because of these issues, it is not completely clear whether general interacting field theories have a null hypersurface formulation.  Some arguments for and against will be given in section \ref{nonpert}.

The plan of this article is as follows: Section \ref{arg} will outline the physical assumptions used to prove the GSL, and show why the GSL follows from them.  Section \ref{scalar} will describe in detail the null hypersurface formulation for a free scalar field.  Section \ref{spin} will generalize these results to free spinors, photons, and gravitons.  Section \ref{int} will discuss what happens when interactions are included.

Conventions: The metric signature will be plus for space and minus for time.  On the horizon, $y$ is a system of $D-2$ transverse coordinates which is constant on each horizon generator, $\lambda$ is an affine parameter on each horizon generator, and $k^a$ points along each horizon generator and satisfies $k^a \nabla_a \lambda = 1$.  When moving off the horizon, $u$ will be a null coordinate such that the horizon is located at $u = 0$, and $v$ will be a null coordinate which satisfies $v = \lambda$ on the horizon, such that the metric on the horizon is
\begin{equation}
ds^2 = -du\,dv + h_{ij} dy^i dy^j.
\end{equation}

To reduce clutter, I will use the notation $v^a X_a \equiv X_v$.

\section{Argument for the GSL}\label{arg}

\subsection{Outline of Assumptions}\label{outline}

In order to prove the GSL, I need to make three basic physical assumptions:
\begin{enumerate}
\item \textbf{Semiclassical Einstein Gravity.}
The proof will apply to the semiclassical regime (section \ref{sreg}), in which all physical effects can be controlled by an expansion in $\hbar G / \lambda^2$, where $\lambda$ is the characteristic de Broglie wavelength of the matter fields.  This expansion is valid when $\lambda \gg L_\mathrm{planck}$.  By holding $\lambda$ and $G$ fixed, one can regard this as an expansion in $\hbar$.  The leading order physics is given by quantum field theory on a fixed classical spacetime.  However, at higher orders in $\hbar$ there are perturbations to the spacetime metric due to gravitational back-reaction.

These perturbations affect the horizon area $A$ at $\mathcal{O}(\hbar^1)$, and therefore affect $S_\mathrm{H}$ at $\mathcal{O}(\hbar^0)$.  At this order, the gravitational backreaction will be treated as a c-number, and will be calculated using the semiclassical Einstein equation $G_{ab} = 8\pi G \langle T_{ab} \rangle$.  It will also be assumed that the matter is minimally coupled to the metric.

\item \textbf{The Existence of a Null Hyperspace Formalism.}
Ignoring the backreaction, matter is described by a quantum field theory on the background spacetime.  The interesting case is when the horizon is stationary (for example, a Killing horizon plus nonstationary matter far from the horizon); otherwise the GSL reduces to the classical Area Law (cf. section \ref{sreg}).  In this case, the QFT which describes matter must have a null hypersurface formulation, i.e. there must be a nontrivial algebra of operators $\mathcal{A}(H)$ corresponding to fields restricted to the horizon itself.

This algebra must satisfy four axioms (section \ref{sym}):  \textit{Determinism} means that all information outside of the horizon can be predicted from the horizon algebra $\mathcal{A}(H)$ together with the algebra $\mathcal{A}(\mathcal{I}^+)$ at future null infinity.  \textit{Ultralocality} means that the fields on different horizon generators are independent, so that the algebra $\mathcal{A}(H)$
tensor-factorizes for spatially-disjoint open subsets in the transverse $y$-directions.\footnote{This is a stronger statement than Microcausality, the assertion that all commutators vanish at spacelike separation.  For example, Ultralocality implies that in the vacuum state, all $n$-point functions of the fields vanish at spacelike separations.  This property may be surprising at first to those familiar with canonical quantization of fields on spacelike surfaces.  However, for free fields on a null surface it obtains because there are no derivatives in the formulae for the null stress-energy $T_{kk}$ or the commutators of fields.}  (Because the fields are distributions it is still necessary to smear them in the transverse directions to obtain well-defined operators.) \textit{Local Lorentz Symmetry} means that the degrees of freedom on each horizon generator are symmetric under translations and boosts.  And \textit{Stability} is the requirement that the fields on each horizon generator have positive energy with respect to the null translation symmetry.  (These four axioms will be explicitly shown for free QFT's in section \ref{scalar}-\ref{spin}.)

In the case of a free field $\phi$, this algebra can contain operators that depend on the pullback of $\phi$ to the horizon $\phi(u = 0)$, but not on e.g. the derivative moving away from the horizon $\nabla_u \phi(u = 0)$.  For this definition, all four axioms will be shown to hold for fields with various spins (sections \ref{scalar}-\ref{spin}).  But in the case of interacting fields, it is not clear which operator(s) should be regarded as the fundamental field.  In this case it will simply be taken as an assumption that there exists some algebra $\mathcal{A}(H)$ satisfying these properties.  Some tentative arguments for and against this assumption will be discussed in section \ref{int}.

\item \textbf{A Renormalization Scheme for the Generalized Entropy.}  Because the entanglement entropy outside of the horizon diverges, any proof that generalized entropy increases must be formal unless this divergence is regulated and renormalized.  Rather than specify a particular renormalization scheme, I will simply describe what properties the scheme must have (section \ref{ren}).  The proof of the GSL depends on proving that the free boost energy $K - TS$ cannot increase as time passes.  Formally, this quantity can be divided into two parts: the boost energy $K$ and the entropy $S$.  Although $K - TS$ can be rigorously defined and is finite, both $K$ and $S$ suffer divergences which must be renormalized.  It is necessary to assume that, when $K$ is written in terms of the renormalized stress-energy tensor, and $S$ is written in terms of the renormalized entropy, the expected relationship between these three quantities continues to hold.  Since this property can be rigorously shown for infinite lattice spin systems \cite{AS77}, it is reasonable to believe that it also holds for quantum field theories.
\end{enumerate}

\noindent In the remainder of this section, the consequences of these three assumptions will be described in more detail.  One of these consequences is that the horizon is thermal with respect to dilations about any slice \ref{thermal}.  This---together with an information theory result known as the ``monotonicity of the relative entropy'' (section \ref{relative})---implies the GSL (sections \ref{proofon}-\ref{outside}).

\subsection{The Semiclassical Regime}\label{sreg}

In the semiclassical approximation, we add certain quantum fields $\phi$ to the classical spacetime, and use their expected stress-energy $\langle T_{ab} \rangle$ as a source for an order $\hbar$ perturbation to the metric.  In the semiclassical limit one takes $\hbar$ to be small, so that the perturbation to the metric is small compared to the classical metric.\footnote{The semiclassical $\hbar$ regime invoked here should be distinguished from the large $N$ semiclassical regime in which one has a large number of particle species and takes $\hbar \to 0$ while holding $\hbar N$ fixed.  In that kind of semiclassical regime the quantum corrections to the metric can be of the same order as the classical metric, so that it is not possible to regard it as a small perturbation.  Proving the GSL in the large $N$ regime will be left for another day.}

The perturbed metric can be expanded in $\hbar$ as:
\begin{equation}
g_{ab} = g_{ab}^0 + g_{ab}^{1/2} + g_{ab}^1 + \mathcal{O}(\hbar^{3/2}).
\end{equation}
The zeroth order term is the classical background metric, the half order term is due to quantized graviton fluctuations, and the first order term is due to the gravitational field of matter or gravitons.  Since the GSL is an inequality, in the limit of $\hbar \to 0$, the truth or falsity of the GSL is determined solely based on the highest order in $\hbar$ contribution to the time derivative of the generalized entropy.

The back-reaction of the quantized fields is the $\mathcal{O}(\hbar^1)$ part of the metric, and will be calculated using the semiclassical Einstein equation:
\begin{equation}\label{semi}
G_{ab} = 8\pi G \langle T_{ab} \rangle,
\end{equation}
in which the Einstein tensor $G_{ab}$ is regarded as a c-number, while the stress-energy tensor $T_{ab}$ is a quantum operator.

A few words are in order about the justification of Eq. (\ref{semi}).  In reality, the metric tensor ought to be quantized just as the matter fields are.  When this is done, one should use not the semiclassical Einstein equation, but the full Einstein equation, interpreted as an operator equation.  However, in the linearized weak-field approximation limit, the semiclassical Einstein equation should be recoverable from the operator Einstein equation by taking expectation values of the $\mathcal{O}(\hbar^1)$ part of the metric \cite{10proofs}.  In addition, there should be higher order in $\hbar$ corrections to the Einstein equation, coming from renormalization theory.  However, because this article only treats back-reaction at leading order in $\hbar$, effects which are higher order in $\hbar$ may be neglected.

Hence, because this article uses the semiclassical expansion only when controlled by an $\hbar$ expansion, the results are presumably in correspondence with the full quantum theory.  This regime is much more circumscribed than the ``self-consistent'' semiclassical solutions of e.g. Flanagan and Wald \cite{FW}.  In particular, pathological features such as run-away solutions are outside of the scope of this regime, since they show up only when all orders in $\hbar$ become important.

\paragraph{Semiclassical Expansion of the Raychaudhuri Equation.}

In the strictly classical $\hbar \to 0$ limit, the horizon entropy $S_\mathrm{H} = 1/{4G \hbar}$ of the GSL dominates over the $S_\mathrm{out}$ term.  For any classical manifold with classical fields obeying the null energy condition $T_{kk} = 0$, the area of any future horizon is required to be nondecreasing by Hawking's area increase theorem \cite{hawking71}.  Let $\theta$ be the expansion of the horizon, and $\sigma_{ab}$ the shear.  Then it follows from the convergence property of the Raychaudhuri equation:
\begin{equation}
\nabla_k \theta = -\frac{\theta^2}{D-2} - \sigma_{ab}\sigma^{ab} - R_{kk},
\end{equation}
together with the null-null component of the Einstein equation
\begin{equation}
R_{kk} = 8\pi G\,T_{kk},
\end{equation}
and the absence of any singularities on the horizon itself, that
\begin{equation}
\theta \ge 0.
\end{equation}
Furthermore, if any generator of the horizon has nonvanishing null energy or shear anywhere, the entropy is strictly increasing along that horizon generator prior to that time.  This is the classical area increase theorem.

This classical result can be used to divide the semiclassical GSL into three cases based on the classical $\mathcal{O}(\hbar^0)$ part of the metric.  Either: 1) the horizon is classically growing, 2) it is classically stationary, or 3) it is classically growing up to a certain time $t$, after which it becomes stationary.  In case (1), the zeroth order area increase corresponds to an $\mathcal{O}(\hbar^{-1})$ increase in the generalized entropy, which dominates over all other effects.  Therefore the GSL holds.  In case (2) quantum effects can cause the area to decrease, and therefore it is an interesting question whether the GSL holds or not.  In case (3), the GSL must be true before time $t$, so the only question is whether it holds after $t$.  But the GSL after $t$ makes no reference to anything that occurred before $t$.  Consequently without loss of generality we need consider only case (2), in which the horizon is always classically stationary.  Any violation of the GSL must come from quantum effects, corresponding to order $\hbar^{0}$ contributions to the generalized entropy.\footnote{This article will not consider contributions to the generalized entropy which are higher order in $\hbar$.  In the semiclassical limit, the only way these higher order corrections could violate the GSL is if the GSL is saturated at order $\hbar^0$.  This would require the fields on the horizon to be in a special state for which the time derivative of the generalized entropy is exactly \emph{zero} at order $\hbar^0$.  Probably the only such equilibrium state is the Hartle-Hawking state.  But in this state, the GSL holds to all orders in $\hbar$, by virtue of time translation symmetry.  Thus, the GSL can be expected to hold to all orders in $\hbar$, in the semiclassical regime.  A more interesting question is what happens outside the semiclassical regime, when all orders in $\hbar$ can become equally important.}

Since there is no half-order contribution to $T_{ab}$ or $\sigma_{ab}\sigma^{ab}$, the half order Raychaudhuri equation says
\begin{equation}\label{Rayhalf}
\nabla_k \theta^{1/2} = 0.
\end{equation}
We can now write the first order part of the Raychaudhuri equation as
\begin{equation}\label{Ray1}
\nabla_k \theta^1 = - \langle \sigma_{ab}^{1/2}\sigma^{ab\phantom{i}{1/2}} \rangle - 8\pi \langle T_{kk}^{1} \rangle.
\end{equation}
The $\theta^2$ term is of order $\mathcal{O}(\hbar^{2})$ and is therefore negligible.  If one ignores gravitons, then the shear term $\sigma_{ab}^{1/2}\sigma^{ab\phantom{i}{1/2}}$ can be neglected.  On the other hand, in processes involving gravitons, the shear term must be included (cf. section \ref{grav}).  The easiest way to handle gravitons is to lump the shear squared term in with $T_{kk}$ as a gravitational analogue of the null energy flux.  Below, the stress-energy tensor should be read as including the shear-squared term, thus:
\begin{equation}\label{linRay}
\nabla_k \theta = -8\pi G\,\langle T_{kk} \rangle.
\end{equation}
So when energy falls across the classically stationary horizon, it makes it no longer stationary at order $\hbar^1$.

Let us now calculate the area $A$ of a slice $\Sigma$ cutting the horizon.  A specific slice $\Sigma$ may be defined by specifying the affine parameter $\lambda = \Lambda(y)$ as a function of the horizon generator.  In order to calculate the effects of $T_{kk}$ on the area $A(\Lambda)$ of the slice, we use the relation between the expansion and the area:
\begin{equation}
\theta = \frac{1}{A}\frac{dA}{d\lambda} = A^{-1} \nabla_k A,
\end{equation}
where $A$ is the area of an infinitesimal cross section of the horizon.  This allows the left-hand side of Eq. (\ref{linRay}) to be rewritten as:
\begin{equation}
\nabla_k A^{-1} \nabla_k A = A^{-1} \nabla_k^2 A - A^{-2} (\nabla_k A)^2,
\end{equation}
where the second term can be dropped in the semiclassical approximation because it is nonlinear in $\nabla_k A$.  Thus the linearized Raychaudhuri Eq. (\ref{linRay}) can be rewritten as
\begin{equation}\label{lin2}
\nabla_k^2 A = -8\pi G\,\langle T_{kk} \rangle A.
\end{equation}
After integrating this twice in the $\lambda$ direction, one obtains for the left-hand side of Eq. (\ref{lin2})
\begin{equation}
\int_\Lambda^\infty d\lambda^\prime \int_{\lambda^\prime}^\infty d\lambda\,\nabla_k^2 A(\lambda) =
-\int_\Lambda^\infty d\lambda^\prime\,\nabla_k A(\lambda^\prime) =
A(\Lambda) - A(\infty),
\end{equation}
by using the fundamental theorem of calculus twice, as well as applying the ``teleological'' boundary condition suitable for a future event horizon:
\begin{equation}
\theta(+\infty) = 0.
\end{equation}
Meanwhile, the identical transformation of Eq. (\ref{lin2})'s right-hand side is
\begin{equation}
-8\pi G \int_\Lambda^\infty d\lambda^\prime \int_{\lambda^\prime}^\infty d\lambda\, \langle T_{kk} \rangle A =
-8\pi G \int_\Lambda^\infty \langle T_{kk} \rangle A (\lambda - \Lambda)\,d\lambda.
\end{equation}
The final step is to integrate the infinitesimal areas A in the $D - 2$ transverse $y$-directions.  One obtains the key relationship
\begin{equation}\label{Tint}
A(\Lambda) = A(+\infty)
- 8\pi G \int_\Lambda^\infty \langle T_{kk} \rangle \,(\lambda - \Lambda) \,d\lambda\,d^{D-2}y \equiv 8\pi G\, \langle K(\Lambda) \rangle,
\end{equation}
where the area element has been absorbed into the definition of the transverse integration measure $d^{D-2}y$.

In the next section it will be seen that $K(\Lambda)$ is the generator of a ``boost'' transformation on the horizon about the slice $\Lambda$.  Thus the physical interpretation of Eq. (\ref{Tint}) is that, up to an additive constant, the boost energy $K$ is proportional to the area:
\begin{equation}\label{AB}
A(\Lambda) = C - 8\pi G\, \langle K(\Lambda) \rangle.
\end{equation}
The constant $C$ can be dropped for purposes of the GSL, which is only concerned with area differences.

In the special case where $\Sigma$ is the bifurcation surface of the unperturbed horizon, Eq. (\ref{Tint}) is the `physical processes' version of the first law of black hole thermodynamics \cite{GW01}, while Eq. (\ref{AB}) indicates that the horizon area is canonically conjugate to the Killing time \cite{CT93}.  But to show the GSL, it is important that these formulae hold even when $\Sigma$ is not the bifurcation surface.

\subsection{Properties of the Horizon Algebra}\label{sym}

As stated above, we are assuming that our matter quantum field theory has a valid null-hypersurface initial- value formalism.  That means that there must be a field algebra $\mathcal{A}(H)$ which can be defined on any stationary horizon $H$ without making reference to anything outside of $H$.  More precisely, all properties of the algebra must be defined using no more than:
\begin{enumerate}
\item some set of quantum operators defined as a local net of algebras over $H$ (e.g. based on local quantum fields which make sense as operator-valued distributions $\phi(\lambda, y)$ on $H$; this will done for free fields in  sections \ref{scalar}-\ref{spin}),
\item the transverse components of the metric $g_{ij}$ on $H$, and
\item an affine parameter $\lambda$ on each horizon generator (which actually depends on a Christoffel symbol $\Gamma^v_{vv} = g_{uv,v}$ in null coordinates).
\end{enumerate}
Assuming that an algebra can be so defined, one expects it to obey the four axioms: Determinism, Ultralocality, Local Lorentz Symmetry, and Stability.  These axioms will be shown in sections \ref{scalar}-\ref{spin} for free fields, but plausibly hold even for interacting fields, assuming that a null hypersurface restriction makes sense at all for such fields (cf. section \ref{int}).

The axiom of Determinism says that $\mathcal{A}(H)$ gives a complete specification of all information falling across the horizon, so that together with the information in $\mathcal{A}(\mathcal{I}^+)$ at null infinity, one can determine all the information outside the event horizon (Fig. 1a).  Consequently, any symmetries of the horizon $H$ will correspond to hidden symmetries of the theory on the bulk.  Thus by working out the symmetry group of $\mathcal{A}(H)$, hidden properties of the bulk dynamics will become manifest.

\begin{figure}[ht]
\centering
\includegraphics[width=.85\textwidth]{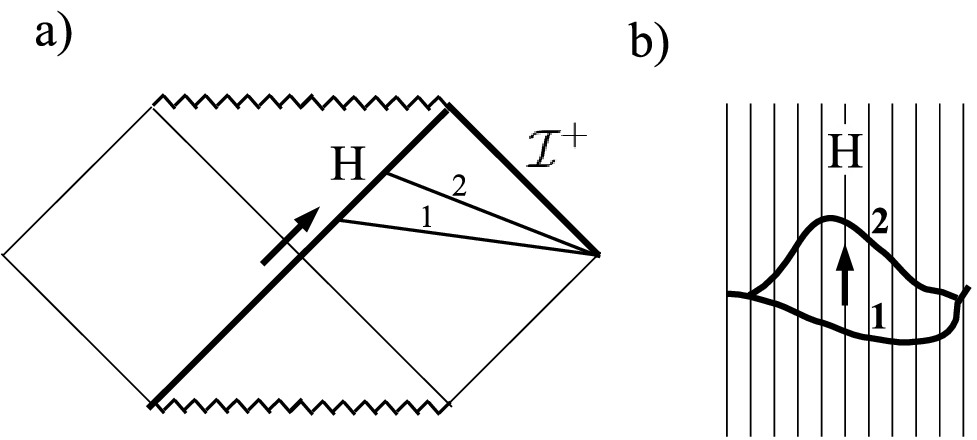}
\caption{\footnotesize a) An eternal black hole spacetime is shown.  The GSL says that the generalized entropy must increase from time slice $1$ to time slice $2$.  However, all information outside of the horizon must either fall across the horizon $H$ or else reach future null infinity $\mathcal{I}^+$ (Determinism).  Hence one can ``push forward'' each of the two time slices to part of $H$ and all of $\mathcal{I}^+$ without losing any information.  In addition to the Killing symmetry which acts on the horizon as a dilation, there is a translation symmetry of $H$ (shown as an arrow) which is not a symmetry of the whole spacetime.
b) A transverse view of $H$ in the same spacetime.  Vertical lines represent horizon generators.  Each horizon generator can be independently translated and dilated (Local Lorentz Symmetry); this permits any two slices on $H$ to be translated into each other, and ensures that region above each slice on $H$ is thermal with respect to dilations about that slice.  In order to prove the GSL this thermal property is needed for both slice $1$ and slice $2$.
} \label{nullslice}
\end{figure}

The axiom of Ultralocality says that the degrees of freedom on different horizon generators are independent systems.  So if the set of horizon generators is written as a disjoint union $H = \sum_n H_n$ of open regions in the transverse $y$-space, then the algebra can be written as a tensor product $\mathcal{A}(H) = \prod_n \mathcal{A}(H_n)$, where $\mathcal{A}(H_n)$ is the algebra of fields restricted to $H_n$.

Ultralocality is stronger than Microcausality, which merely asserts that the commutators of fields vanish at spacelike separation.  In particular, in the vacuum state (whose existence is guaranteed by the other axioms of Local Lorentz Invariance and Stability), Ultralocality implies that all $n$-point functions of field operators in $\mathcal{A}(H)$ vanish when evaluated on $n$ distinct horizon generators.

This property may be shocking to those who are used to canonical quantization on a spacelike initial data surface, because on a spacelike surface it is impossible for any Hadamard state to have vanishing entanglement across short spatial distances.  By contrast, on a stationary null surface the vacuum entanglement is arranged solely along each horizon generator and not between different horizon generators \cite{schroer09}.  In the case of a free bosonic field $\phi$, this vanishing of $n$-point functions is possible because 1) the null stress-energy $T_{kk}$ does not depend on transverse $y$-derivatives of the field, but only the null derivative $\nabla_k \phi$, and 2) the horizon algebra $\mathcal{A}(H)$ does not include the field $\phi$ itself (which has nonvanishing $n$-point functions at spacelike separation on the null surface), but only $\nabla_k \phi$ (which does not).\footnote{A nice exercise is to demostrate explicitly, for a free massive scalar $\Phi$ in Minkowski space, in a null coordinate system $(u, v, y_i)$ that the $n$-point functions of $\nabla_v \Phi$ vanish when evaluated on the null surface $u = 0$ for distinct values of $y$.}

Because Ultralocality requires that the different horizon generators can be treated as independent systems (although the field operators still need to be integrated in the transverse $y$-directions to give well-defined operators), the remaining two axioms, Lorentz Symmetry and Stability, can without loss of generality be applied to each horizon generator separately.

Local Lorentz Symmetry means that the algebra $\mathcal{A}(H)$ has an infinite dimensional group $G$ of symmetries (i.e. automorphisms) corresponding to affine transformations of each horizon generator:
\begin{equation}
\delta \lambda = a(y) + b(y) \lambda,
\end{equation}
$a$ and $b$ being functions of $y$.  Each horizon field $\phi(\lambda, y)$ must transform in some representation of this group (just as fields in flat spacetime transform in some representation of the Poincar\'{e} group).  Thus, each element $g \in G$ has both a geometrical interpretation (acting on horizon points) and an operator interpetation (acting on fields).  Compatibility of these two interpretations requires that if an operator $\mathcal{O}$ is localized in a region $R$, then $g(\mathcal{O})$ is localized in $g(R)$.

This is quite a bit more symmetry than can be possessed by the spacetime in which $H$ is embedded (Fig 1b).  These secret symmetries of $H$, together with the other assumptions, will turn out to imply the GSL.  (In the case of free fields, it will be shown in section \ref{conf} that the horizon algebra is also invariant under special conformal transformations $\delta \lambda = c(y) \lambda^2$, but this additional symmetry is not required to prove the GSL.)

In order to implement these symmetries, we need not only the field $\phi$, but also certain integrals of the $T_{kk}$ component of the stress-energy tensor.  This component of the stress-energy tensor represents the flux of null energy across the horizon.  Since the null energy is the generator of null diffeomorphisms, $T_{kk}$ can be integrated to obtain the generator of affine reparameterizations.

The generator of a null translation $\delta \lambda = a(y)$ is given by
\begin{equation}\label{pka}
p_k(a) \equiv \int T_{kk}\,d\lambda\,a(y)\,\,d^{D-2}y.
\end{equation}
(Here and below, the area element of the horizon will be considered to be implicit in the integration measure $d^{D-2}y$.)

Stability says that so long as $a(y) > 0$, $p_k \ge 0$.  In other words, the generator of null translations must be nonnegative.  By taking the limit in which the amount of translation is a delta function ($a(y) \to \delta^{D-2}(y)$), one finds that Stability is equivalent to the average null energy condition (ANEC) \cite{borde87}, as evaluated on horizon generators;
\begin{equation}
p_k(y) \equiv \int_{-\infty}^{+\infty} T_{kk}\,d\lambda.
\end{equation}
The ANEC is a manifestation of the positivity of energies in a quantum field theory.\footnote{The ANEC can be derived from the stability of the quantum field theory by the following argument: any stationary horizon $H$ can be embedded in a spacetime $\mathcal{M}_{1,1} \otimes (\Sigma \cap H)$, where the first factor is 1+1 dimensional Minkowski space, and the second is some $D-2$ dimensional Riemannian manifold.  Now suppose that the quantum fields have their energy bounded below, relative to time translation on $\mathcal{M}_{1,1}$.  By Lorentz symmetry and continuity, the null energy on $\mathcal{M}_{1,1}$ must also be bounded below.  All null energy must eventually cross the horizon $H$, hence the null energy on $H$ is bounded below.  But by Ultralocality this is only possible if each horizon generator is separately stable.}   It is possible to show that the ANEC holds on the null generators of a stationary horizon by invoking the GSL \cite{anec}.  Here we go in the converse direction, using the ANEC to help prove the GSL.

Given any $a(y) > 0$, it is possible to define the vacuum state $|0\rangle$ on the horizon as being the ground state with respect to the null energy $p_k(a)$ \cite{sewell82}.  However, in an ultralocal theory, there can be no interaction between the different horizon generators.  Therefore the state factorizes: it is a ground state with respect to each $p_k(y)$ separately.  This means that each possible choice of $a(y) > 0$ defines the \emph{same} vacuum state.

We can also perform a dilation $\delta \lambda = b(y)\lambda$.  This symmetry is generated by
\begin{equation}
K(y) \equiv \int T_{kk}\,\lambda\,d\lambda\,b(y)\,d^{D-2}y.
\end{equation}
For any particular spatial slice of the horizon located at $\lambda = \Lambda(y)$, one can define a canonical `boost energy' $K$ of the horizon in the region $\lambda > \Lambda(y)$:
\begin{equation}\label{modular}
K(\Lambda) \equiv \int_\Lambda^\infty T_{kk}\,(\lambda - \Lambda) \,d\lambda\,d^{D-2}y.
\end{equation}
The definition of $K$ depends on the slice $\Lambda(y)$ in two different ways: not only does the lower limit of integration change, but the horizon Killing vector $\lambda - \Lambda$ which preserves the slice $\Lambda$ also changes.  The next section will show that the vacuum state $|0\rangle$ is thermal in the region $\lambda > \Lambda(y)$ with respect to $K(\lambda)$, no matter what slice $\Lambda$ is chosen (Fig. 1b).

\subsection{Thermality of the Horizon}\label{thermal}

The purpose of this section is to show that $|0\rangle$ is thermal with respect to boosts when evaluated above any arbitrary slice $\Lambda$ on the horizon.  The boost acts geometrically on each horizon generator $y$:
\begin{equation}
(\lambda - \Lambda(y)) \to e^{t} (\lambda - \Lambda(y)).
\end{equation}
The axiom of Local Lorentz Invariance requires that this geometrical action of the boost correspond to an automorphism of the algebra of observables $\mathcal{A}(\lambda > \Lambda)$ localized above the slice $\Lambda$.  This induces a 1-parameter group of automorphisms $\alpha_t$ acting on operatos in $\mathcal{A}(\lambda > \Lambda)$.

\paragraph{KMS States.}  The thermality of the vacuum state $|0\rangle$ means that it obeys the Kubo-Martin-Schwinger (KMS) condition: For any two observables $A$ and $B$, $\langle B \alpha_{t}(A) \rangle_0$ must be an analytic function of $z$ when $0 < \textrm{Im}(t) < i \hbar \beta$, and also
\begin{equation}\label{KMS}
\langle AB \rangle_0 = \langle B\,\alpha_{i \hbar \beta}(A) \rangle_0,
\end{equation}
where $\beta = 2\pi / \hbar$ is the inverse Unruh temperature.  In order to establish this, we appeal to an analogue of the Bisognano-Wichmann theorem.

The Bisongano-Wichmann theorem \cite{BW76} implies that for any set of quantum fields in Minkowski space (interacting or not) satisfying the Wightman axioms, in the vacuum state $|0\rangle$, the fields restricted to a Rindler wedge $W$ are thermal with respect to the boost energy.  This is the Unruh effect.  The basic inputs of the theorem are 1) the Lorentz symmetry of the wedge, and 2) the spectral condition (i.e. positivity of energies) with respect to time translation.

The basic idea of their (highly technical) theorem is to analytically continue the boost symmetry of the group to complex values.  One can then boost a Rindler wedge by an amount $i\pi$ in order to ``rotate'' it into the complementary Rindler wedge region $W^\prime$ on the other side of the bifurcation surface.  This rotation corresponds to acting with the operator $e^{\pi K /\hbar}$, where $K$ is the generator of the boost symmetry in $W$.

Using the spectral condition to ensure convergence, Bisognano and Wichmann showed that each operator $\mathcal{O}$ in the wedge algebra $\mathcal{A}(W)$ satisfies
\begin{equation}\label{BW}
J e^{-\pi K / \hbar} \mathcal{O} |0\rangle = \mathcal{O}^* |0\rangle,
\end{equation}
where $J$ is the (antiunitary) CPT symmetry transformation corresponding to reflecting one time and one space dimension through the bifurcation surface of $W$, and $*$ is hermitian conjugation.  Sewell \cite{sewell80} observed that Eq. (\ref{BW}) implies that $|0\rangle$ is a KMS state with temperature $\hbar / 2\pi$, with respect to boosts, when restricted to $W$.  This is because
\begin{eqnarray}
\langle 0| AB |0\rangle = \langle 0| A^* e^{-\pi K/\hbar} J \cdot J e^{-\pi K/\hbar} B^* |0\rangle = \\
\langle 0| B e^{-2\pi K/\hbar} A |0\rangle = \langle 0| B\,\alpha_{2i\pi}(A) |0\rangle,
\end{eqnarray}
where in going from the first line to the second we have used the fact that $J$, being antiunitary, converts bras to kets and vice versa.  Sewell also observed that, given certain axioms, the Bisognano-Wichmann theorem could be applied to black hole spacetimes to derive thermality outside of the bifurcation surface of a black hole.

In a later article \cite{sewell82}, Sewell generalized the Bisongano-Wichmann theorem further to the case of a quantum field algebra restricted to a stationary horizon (under the assumption that this algebra exists).  In this generalization, 1) the dilation symmetry $b$ is analogous to the boost symmetry, and 2) Stability with respect to translation symmetry $a$ is analogous to the spectral condition.  This generalization can be used here to show that when the vacuum state $|0\rangle$ is restricted to the region $\lambda > \Lambda$, it is a KMS state with respect to the dilation generated by $K(\Lambda)$, with a temperature $T = \hbar / 2\pi$.  This is just the Unruh/Hawking effect as viewed on the horizon itself.

In Sewell's construction, $|0\rangle$ is simply the Hartle-Hawking state associated with the fields on the horizon $H$ itself.  This means that if the bulk spacetime possesses a Hartle-Hawking state, it will restrict to $|0\rangle$ on $H$.  However, even in spacetimes which do not possess a Hartle-Hawking state (such as the Kerr black hole), the state $|0\rangle$ is still well-defined.  This fills a lacuna in certain previous proofs of the GSL, which did not apply to such horizons \cite{10proofs}.

For an algebraic proof that the vacuum of any chiral CFT obeys the thermality property, see \cite{gabbiani1993}.


\paragraph{Gibbs States.} Another definition of thermal states which is sometimes used is the Gibbs definition, in which a thermal state with respect to some Hamiltonian (in this case the boost energy $K$) is defined as the exponentially decaying density matrix
\begin{equation}\label{Gibbs}
\frac{e^{-\beta K}}{\mathrm{tr}(e^{-\beta K})},
\end{equation}
where the denominator is the partition function.  The relationship between the KMS and Gibbs definitions is as follows: in situations with a finite number of degrees of freedom, in which the algebra of observables $\mathcal{A}$ is just type I (i.e. the complete collection of operators on a Hilbert Space), the Gibbs and the KMS definitions are equivalent.\footnote{The standard way to show this is to plug Eq. (\ref{Gibbs}) into Eq. (\ref{KMS}), and use the cyclic property of the trace.}  However, in QFT there are an infinite number of degrees of freedom, and typically the resulting algebras are type III (meaning that there is no trace operation).  In this case, the KMS definition still works, while the Gibbs definition becomes ill-defined.  Nevertheless, it is a common practice in QFT to formally manipulate expressions like Eq. (\ref{Gibbs}) in order to extract finite answers.  Such procedures can in principle be justified by a renormalization procedure in which one regulates the divergences in Eq. (\ref{Gibbs}) and then renormalizes.

Using this less rigorous Gibbs definition, the thermality of the Rindler wedge can also be proven by a simple path integral argument developed by Unruh and Weiss \cite{UW84}.  Assuming that the vacuum state $|0\rangle$ is the lowest energy state, it can be generated by the boundary of a Euclidean path integral extending from time $t = -\infty$ to $t = 0$.  Expressed in terms of the Hamiltonian $H$ and the partition function $Z$,
\begin{equation}
|0\rangle = \lim_{t \to \infty} \frac{e^{- tH / \hbar}}{\mathrm{tr}(e^{- tH / \hbar})}.
\end{equation}
However, this same Euclidean path integral can be viewed in radial coordinates as a path integral extending from an angle $\theta = 0$ to an angle $\theta = \pi$.  This indicates that when one traces out over the degrees of freedom in the complementary wedge $W^\prime$, the state of $W$ is
\begin{equation}
\sigma = \frac{e^{-2 \pi K/\hbar}}{\mathrm{tr}(e^{- 2\pi H / \hbar})},
\end{equation}
which is thermal.

In order to show an analogous Unruh-Wiess thermality for the horizon algbera, one would have to find a way to write the vacuum state $|0\rangle$ in terms of a path integral over a complexified $\lambda$ coordinate.  The periodicity of the path integral in radial coordinates would then imply the thermality of the restricted vacuum with respect to the boosts.  However, it is not entirely clear what the conditions are for such a path integral to exist.  In sections \ref{scalar} and \ref{spin}, it will be shown how to reduce free fields restricted to the horizon to free left-moving conformal fields in 1+1 dimension, which would allow the vacuum state to be written in terms of free two-dimensional path integrals.

In conclusion, there exist proofs of the thermality of the vacuum in the Wightman, algebraic, and path integral approaches to QFT.  The first two approaches prove that the vacuum is thermal in the KMS sense, while the third is a formal demonstration using the less rigorous Gibbs definition.  All three approaches are potentially capable of being adapted to the observables living on the horizon itself.  However, the algebraic approach currently assumes special conformal symmetry, and the path integral approach must of course assume the existence of a path integral.

\subsection{The Relative Entropy}\label{relative}

In order to prove that the generalized entropy increases, I need to use a monotonicity property of an information-theoretical quantity known as the ``relative entropy''.  The relationship between the relative and generalized entropies was made explicit in Casini \cite{casini08}, and was used in my earlier proof of the GSL for Rindler wedges \cite{rindler}.

For a finite dimensional system, the relative entropy of two states $\rho$ and $\sigma$ is defined as
\begin{equation}\label{relIII}
S(\rho\,|\,\sigma) = \mathrm{tr}(\rho\,\ln\,\rho) - \mathrm{tr}(\rho\,\ln\,\sigma).
\end{equation}
For a QFT system with infinitely many degrees of freedom, it may be defined as the limit of this expression as the number of degrees of freedom go to infinity \cite{araki75}.\footnote{The von Neumann algebra of a bounded region in a QFT is a hyperfinite type III algebra \cite{BAF87}.  Hyperfinite means that one can approximate it by a series of finite dimensional algebras; hence the limit.  Because of the monotonicity property, it does not matter how the limit is taken.}  The relative entropy lies in the range $[0,\,+\infty]$.  In some sense it measures how far apart the two states $\rho$ and $\sigma$ are, but it is asymmetric: $S(\rho\,|\,\sigma)$ is not in general the same as $S(\sigma\,|\,\rho)$.

\paragraph{Examples} When the two states are the same the relative entropy vanishes:
\begin{equation}
S(\rho\,|\,\rho) = 0.
\end{equation}
When $\sigma = \Psi$ is a pure state and $\rho \ne \Psi$, the relative entropy is infinite:
\begin{equation}
S(\rho\,|\,\Psi) = +\infty.
\end{equation}
Normally, one wants to use a faithful state for $\sigma$ (i.e. one without probability zeros) so that $S(\rho\,|\,\sigma)$ is finite on a dense subspace of the possible choices for $\rho$.

When $\sigma$ is the maximally mixed state in an $N$ state system, the relative entropy is just the entropy difference:
\begin{equation}
S(\rho\,|\,1/N) = \ln N - S_\rho.
\end{equation}

When $\sigma$ is a Gibbs state with respect to a some Hamiltonian `energy' $H$, the relative entropy $S (\rho\,|\,\sigma)$ is the difference of free energy, divided by the temperature:
\begin{equation}\label{FE}
S(\rho\,|\,\sigma) = [(\langle H \rangle_\rho - T_\sigma S_\rho) - (\langle H \rangle_\sigma - T_\sigma S_\sigma)]/T_\sigma,
\end{equation}
where $T_\sigma$ is the temperature of $\sigma$.  This can be verified by inserting Eq. (\ref{Gibbs}) into Eq. (\ref{relIII}).

One would also like to be able to apply Eq. (\ref{FE}) to KMS states of systems with infinitely many degrees of freedom, even when the Gibbs definition of thermality is ill-defined.\footnote{In fact, all faithful states can be regarded as KMS states with respect to some notion of `time' defined relative to that state \cite{summers05}.  This notion of time evolution is known as the ``modular flow''.}  Although the relative entropy itself is typically finite for sufficiently reasonable states, the individual components $H$ and $S$ can diverge.  The GSL proof presented in the next section assumes that Eq. (\ref {FE}) can be applied even in this context so long as one uses the \emph{renormalized} entropy and energy values.  Some evidence for this unproven assumption will be discussed in section \ref{ren}.

\paragraph{Monotonicity} However, the most important property of the relative entropy is that it monotonically decreases under restriction.  Given any two mixed states $\rho$ and $\sigma$ defined for a system with algebra $M$, if we restrict to a smaller system described by a subalgebra of observables $M^\prime$, the relative entropy cannot increase \cite{lindblad75}:
\begin{equation}
S(\rho\,|\,\sigma)_M \ge S(\rho\,|\,\sigma)_{M^\prime}.
\end{equation}
Intuitively, since the relative entropy measures how different $\rho$ is from $\sigma$, if there are less observables which can be used to distinguish the two states, the relative entropy should be smaller.

\subsection{Proving the GSL on the Horizon}\label{proofon}

The monotonicity property looks very similar to the GSL.  And in fact, with the right choice of $\rho$ and $\sigma$ it is the GSL.

It was observed in section \ref{thermal} that the vacuum state $|0\rangle$ defined on $H$ is a KMS state with respect to $K(\Lambda)$, no matter what $\Lambda$ slice is chosen.  Therefore, under horizon evolution a thermal state restricts to another thermal state.  Of course, the GSL holds trivially for this vacuum state $|0\rangle$ because of null translation symmetry---the goal is to prove it for some other arbitrary mixed state of the horizon.  Let $\rho(H)$ be the state of the horizon algebra $\mathcal{A}(H)$ which we wish to prove the GSL for, and let $\sigma = | 0 \rangle \langle 0 |$ be the vacuum state with respect to null translations.

Since $\sigma$ is a KMS state when restricted to the region above any slice, the relative entropy $S(\rho\,|\,\sigma)$ is a free energy difference of the form Eq. (\ref{FE}), where $E$ is the boost energy $K(\Lambda)$ of the region $\lambda > \Lambda$, $S$ is the entropy of $\lambda > \Lambda$, and $T = \hbar/{2\pi}$ is the Unruh temperature.

Furthermore by virtue of null translation symmetry, $(\langle K \rangle - TS)_\sigma$ is just a constant.  So the monotonicity of relative entropy therefore tells us that as we evolve from a slice $\Lambda$ to a later slice $\Lambda^\prime$,
\begin{equation}
\frac{2\pi}{\hbar} \langle K(\Lambda) \rangle - S(\Lambda) \ge
\frac{2\pi}{\hbar} \langle K(\Lambda^\prime) \rangle - S(\Lambda^\prime),
\end{equation}
Using Eq. (\ref{AB}), this implies that the GSL holds on the horizon for the state $\rho(H)$:
\begin{equation}
\frac{A}{4\hbar G}(\Lambda^\prime) + S(\Lambda^\prime) \ge \frac{A}{4\hbar G}(\Lambda) + S(\Lambda).
\end{equation}

\subsection{The Region Outside the Horizon}\label{outside}

This does not yet amount to a complete proof of the GSL, because the GSL refers to the entropy $S_\mathrm{out}$ on a spacelike surface $\Sigma$ \emph{outside} of $H$, not just to the entropy which falls across $H$.  Depending on how $H$ is embedded in the spacetime, it cannot necessarily be assumed that all of the information on $\Sigma$ will fall across the horizon.  Some of it may escape.\footnote{One might wonder how it is possible for the GSL to hold both on the horizon and outside the horizon, considering that for an evaportating black hole, the generalized entropy only increases due to counting the entropy of Hawking radiation that escapes from the black hole.

The resolution involves the role of the UV cutoff on entanglement entropy, discussed in section \ref{ren}.  The proof of the GSL on the horizon involves choosing the UV cutoff to be at a fixed affine parameter distance $\Delta \lambda$ from each slice.  On the other hand, the GSL outside the horizon requires the UV cutoff to be at a fixed proper distance $\Delta x = \Delta \lambda \Delta u$, where $u$ is the other null coordinate.  Both cutoffs can be simultaneously implemented by choosing $\Delta u$ to be covariantly constant with respect to $\lambda$, but in that case $\Delta u$ is exponentially growing with respect to the radial coordinate $r$, and as a result no Hawking radiation escapes past $\Delta u$.  Alternatively, one could boost the cutoff $\Delta x$ so that it is invariant under the Killing symmetry, but then $\Delta \lambda$ is no longer constant, so the horizon GSL no longer applies.  However, $S_\mathrm{out}$ is unaffected, so the outside GSL remains valid.}

Suppose we have an arbitrary quantum state $\rho$ defined on the region of spacetime $R$ exterior to some stationary horizon $H$.  All of the information in $R$ should either fall across the horizon $H$ or else escape to future infinity $\mathcal{I}^+$.  (This assumes that any singularities are hidden behind $H$---otherwise the information falling into these will need to be included as well.)  $H$ and $\mathcal{I}^+$ should factorize into independent Hilbert spaces, but $\rho$ may be some entangled state on $H\,\cup\,\mathcal{I}^+$.

We can now generalize the proof above by choosing a reference state $\sigma$ that factors into the vacuum state on $H$ times some other state:
\begin{equation}
\sigma(H\,\cup\,\mathcal{I}^+) = |0\rangle \langle 0|(H) \otimes \sigma(\mathcal{I}^+).
\end{equation}
The second factor $\sigma(\mathcal{I}^+)$ can be chosen to be any faithful state (so long as the relative entropy $S(\rho\,|\,\sigma)$ is finite).  After slicing the horizon at $\Lambda(y)$, the relative entropy is then once again a free energy with respect to some modular energy $E$:
\begin{equation}
S(\rho\,|\,\sigma) = ( \langle E \rangle - S)_\rho - (\langle E \rangle - S)_\sigma,
\end{equation}
where because $\sigma$ is a product state, the modular energy $E$ is a sum of terms for the horizon system $H_{\lambda > \Lambda}$ and $\mathcal{I}^+$:
\begin{equation}
E(H_{\lambda > \Lambda}\,\cup\,\mathcal{I}^+) = \frac{2\pi}{\hbar} K(\Lambda) + E(\mathcal{I}^+),
\end{equation}
with $E(\mathcal{I}^+)$ being the modular energy conjugate to the modular flow of $\sigma(\mathcal{I}^+)$.  The addition of the new modular energy term $E(\mathcal{I}^+)$ makes no difference to $\Delta E$, the change in the relative entropy with time, because $E(\mathcal{I}^+)_\rho$ is not a function of the horizon slice $\Lambda$.  Consequently one can still use Eq. (\ref{AB}) to show that
\begin{equation}
\langle \Delta E \rangle = \frac{2\pi}{\hbar} \langle \Delta K \rangle = -\frac{\Delta A}{4\hbar G}.
\end{equation}
On the other hand, $S$ is now interpreted as the total entropy of $\rho$ on on the combined system $H_{\lambda > \Lambda}\,\cup\,\mathcal{I}^+$.  Because of unitarity, the entropy $S(\Sigma)$ of any slice $\Sigma$ that intersects the horizon at $\Lambda$ must be the same as the entropy $S(H_{\lambda > \Lambda}\,\cup\,\mathcal{I}^+)$.  In other words, $S = S_\mathrm{out}$, for any state $\rho$.  (Note that $\rho$, unlike $\sigma$, may have entanglement between $H$ and $\mathcal{I}^+$.)  Thus, the monotonicity property of $S(\rho\,|\,\sigma)$ is equivalent to the GSL.

\subsection{Renormalization}\label{ren}

It should be noted that in every QFT, $K$ and $S$ are both subject to divergences.  The relative entropy packages all of these divergent quantities together in a way that can be rigorously defined for arbitrary algebras of observables \cite{araki75}.  However, in order to apply the Raychaudhuri equation (as needed to obtain Eq. (\ref{AB})) it is necessary to unpackage the relative entropy into separate $K$ and $S$ terms, each of which needs to be renormalized separately.  Because of the connection between the relative entropy and the free energy for finite dimensional subsytems, one expects that after defining $K$ in terms of the renormalized stress-energy tensor $\tilde{T}_{kk}$, and the entropy in terms of some renormalized entropy $\tilde{S},$\footnote{The proper way to renormalize the entropy is not completely clear, but one promising regulator scheme uses the ``mutual information'' between two regions at finite spatial separation \cite{casini06}.} that Eq. (\ref{FE}) still holds:
\begin{equation}
S(\rho\,|\,\sigma) = [(\langle \tilde{K} \rangle - T\tilde{S})_\rho -
(\langle \tilde{K} \rangle - T \tilde{S})_\sigma]/T.
\end{equation}
This is especially plausible given that the only quantities that enter into Eq. (\ref{FE}) are energy and entropy \emph{differences}.

As in my previous proof for Rindler horizons \cite{rindler}, I will assume that this equation is in fact true in an appropriate renormalization scheme.  There is a theorem to this effect for quantum spin systems on an infinite lattice \cite{AS77}, and it seems likely that any QFT can be approximated arbitrarily well by such a lattice.

If one wishes to interpret the GSL as a statement about a regulated entanglement entropy on a spacelike surface, then it is also necessary for the regulator scheme defining $\tilde{S}$ on the null surface $H\,\cup\,\mathcal{I}^+$ to give the same answer as the regulator scheme defining $\tilde{S}_\mathrm{out}$ on a spacelike surface $\Sigma$.  This is a plausible assumption since there exist choices of $\Sigma$ which are arbitrarily close to $H$.  But it is not entirely trivial, because the way that the entropy divergence is localized on a null surface is different from the way it is localized on a spacelike surface.

In the case of a spacelike surface the entropy can be regulated by cutting off all entropy closer than a certain distance $x_0$ to the boundary.  As $x_0 \to 0$, the divergence with respect to that cutoff then scales like $x_0^{2-D}$ on dimensional grounds.

This method cannot work on $H$ because there is no invariant notion of distance along the horizon generators.  By dimensional analysis, this means that the entropy must be logarithmically divergent along the null direction.  Therefore, there is an infrared divergence as well as an ultraviolet divergence.

Even if one cuts off the entropy at an affine distance $\lambda_U$ in the ultraviolet and $\lambda_I$ in the infrared, the entanglement entropy is still infinite due to the infinite number of horizon generators.  One must in addition regulate by e.g. discretizing the space of horizon generators to a finite number $N$.  One then finds that the entropy divergence of the vacuum state scales like
\begin{equation}\label{nulldiv}
S_\mathrm{div} \propto N (\ln \lambda_I - \ln \lambda_U).
\end{equation}
(Cf. section \ref{conf} for a justification of this statement.)  The renormalized entropy $\tilde{S}$ can then found by subtracting the entropy of the vacuum state:
\begin{equation}
\tilde{S}(\rho) = S(\rho) - S(\sigma).
\end{equation}

It is reasonable to hope that this renormalized entropy is the same as the renormalized entropy defined on a spatial slice.  Formally, one can simply take the limit of the entropy difference as a spatial slice $\Sigma$ slants closer and closer to $H$.  However, the renormalization of the generalized entropy is itself a limiting process, so there are issues involving orders of limits.  The analysis of section \ref{outside} implicitly assumes that these limits commute.

Another consequence of renormalization is to add higher curvature contributions to the Lagrangian (cf. section \ref{nonmin}) \cite{jacobson94}.  For example, for free fields in 4 dimensional spacetime, the coefficients of the curvature squared terms in the Lagrangian are logarithmically divergent.  This would invalidate the assumption that the matter is minimally coupled to general relativity.  Fortunately, this effect can be neglected here, because the effects of these higher order terms on the generalized entropy are of higher order in $\hbar$.


\section{Quantizing a Free Scalar on the Horizon}\label{scalar}

The proof of the GSL in section \ref{arg} was incomplete: it depended on four axioms describing the properties of quantum fields on the null surface.  The purpose of this section is to explicitly show how these axioms are satisfied in the simplest case: a free scalar field.  This completes the proof in section \ref{arg} of the semiclassical GSL.

Since the reader may not be familiar with the technical issues regarding null quantization, this section will demonstrate null surface quantization for a free, minimally coupled scalar field $\Phi$ with mass $m^2 \ge 0$ in $D > 2$ dimensions.  This is a quick way to construct the algebra of observables $\mathcal{A}(H)$.  It will be shown that this algebra is nontrivial, and obeys the four axioms required to prove the GSL: Determinism, Ultralocality, Local Lorentz Symmetry, and Stability.

It will also be shown that the horizon algebra can be approximated by the left-moving modes in a large number of 1+1 dimensional free conformal field theories.  This allows one to understand, using the conformal anomaly, why the horizon algebra is not symmetric under arbitrary reparameterizations of $\lambda$, but only special conformal transformations.

The discussion of null quantization will be confined mostly to those issues which are of interest in determining the symmetry properties of the horizon.  For a more detailed review of null quantization, including a fuller treatment of the technically difficult ``zero modes'', consult Burkardt \cite{burkardt96}.

\subsection{Stress-Energy Tensor}

The Lagrangian of the Klein-Gordon field is
\begin{equation}
\mathcal{L} = \Phi(\nabla^2 - m^2)\Phi /2.
\end{equation}
The classical stress-energy tensor on the horizon $H$ can be derived by varying with respect to the null-null component of the inverse metric\footnote{A previous version of this article incorrectly included a factor of (1/2) in the formula for $T_{kk}$.  There was a compensating factor of 2 error in the null commutator (\ref{nullcom}).}:
\begin{equation}\label{PhiTkk}
T_{kk} = -2\frac{\mathcal{\delta L}}{\delta g^{ab}} k^a k^b = (\nabla_k \Phi)^2.
\end{equation}
This is positive except when $\Phi$ is constant, and depends only on the pullback of $\Phi$ to $H$.  The total null energy on the horizon can be found by inserting Eq. (\ref{PhiTkk}) into Eq. (\ref{pka}):\footnote{This formula would have to be modified if the scalar field had a nonminimal coupling term $\Phi^2 R$.  However, the horizon entropy $S_\mathrm{H}$ is also modified (cf. section \ref{nonmin}).  The nonminimally coupled theory must also obey the GSL, since it is equivalent to a minimally coupled scalar by a field redefinition \cite{2ndlaw}.}
\begin{equation}\label{nullE}
p_k = \int (\nabla_k \Phi)^2\,d\lambda\,d^{D-2}y.
\end{equation}
The positivity of this quantity indicates that $\mathcal{A}(H)$ satisfies Stability.  Classically this positivity is obvious.  Quantum mechanically, this expression is divergent.  After subtracting off this divergence, one finds that $T_{kk}$ is actually unbounded below.  Nevertheless, the integral of $T_{kk}$ is bounded below by a vacuum state.  This will become obvious after a Fock space quantization is performed in section \ref{fock}.

\subsection{Equation of Motion and Zero Modes}

For the purposes of specifying initial data, $\lambda$ acts more like a space dimension than a time dimension, in the sense that the value of $\Phi$ at one value of $\lambda$ is (almost) independent of the value of $\Phi$ at other values of $\lambda$.  However, there are some zero mode constraints on the field which must be treated carefully.  There are also some convergence properties required if the total flux of momentum across the null surface is to be finite.


The Klein-Gordon equation of motion is
\begin{equation}
(\nabla^2 - m^2)\Phi = 0.
\end{equation}
This equation can be written in terms of horizon coordinates as
\begin{equation}\label{invert}
\nabla_u \Phi = \nabla_v^{-1} (\nabla_y^2 - m^2)\Phi.
\end{equation}
This equation \emph{almost} permits us to arbitrarily specify $\Phi(y,\,\lambda)$ as `initial data' on $H$.  The only constraint is that $\nabla_u \Phi$ must be finite.  This requires that the operator $\nabla_v$ be invertible, which places constraints on the zero modes of $\Phi(\lambda)$.

If one decomposes $\Phi$ into its Fourier modes:
\begin{equation}\label{omega0}
\tilde{\Phi}(y,\,\omega) = \int
\frac{e^{-i\omega \lambda}}{\sqrt{2\pi}} \Phi(y,\,\lambda)\,d\lambda,
\end{equation}
then $\nabla_v^{-1} = 1/\omega$, which is singular at $\omega = 0$.  Thus for Eq. (\ref{invert}) to be well-defined, it is necessary to require that
\begin{equation}\label{zeromode}
\int_{-\infty}^{+\infty} \Phi\,d\lambda = \mathrm{finite}.
\end{equation}
An exception for this arises when $m = 0$, for solutions which are also zero modes in the $y$ direction (i.e. they lie in the kernel of $\nabla_y^2)$.  In this case, Eq. (\ref{invert}) becomes undefined rather than infinite.  Thus one can add a mode defined by
\begin{equation}\label{zero}
\int^{+\infty}_{-\infty} \Phi\,d\lambda = C,
\end{equation}
for some $C$ which is constant over the whole (connected component of) $H$.

In addition to the zero mode constraints, it is natural to require that the flux of stress-energy across the horizon be finite.  In order for the null momentum to be finite, one needs the integral of $T_{kk}$ to converge:
\begin{equation}
\int_{-\infty}^{+\infty} (\nabla_k \Phi)^2 \,d\lambda = \mathrm{finite}.
\end{equation}
One can also demand that the other components of momentum have finite flux over the horizon.  This leads to an additional constraint:
\begin{equation}\label{intphi}
\int_{-\infty}^{+\infty} m^2 \Phi^2 \,d\lambda = \mathrm{finite},
\end{equation}
which is a nontrivial constraint only for a massive field.  This permits massless fields to have soliton-like solutions in which the asymptotic behavior of $\Phi$ at $\lambda = +\infty$ may differ from the behavior at $\lambda = -\infty$.

In the Fourier transformed description, the field should look like this near $\omega = 0$:
\begin{equation}
\tilde{\Phi}(y,\,\omega) = c_1 \delta(0) + \frac{c_2}{\omega} + c_3(y) + \mathcal{O}(\omega),
\end{equation}
where $c_1$ corresponds to constant $\Phi$, $c_2$ corresponds to a soliton with $\Phi(+\infty) = -\Phi(-\infty)$, and $c_3$ corresponds to the value of the integral (\ref{intphi}).  For a massive field, $c_1 = c_2$ = 0.\footnote{Because of the noninvertibility of $\omega = 0$, one might be tempted to require that $c_3 = 0$ as well, but this would be a mistake.  First of all,
$\tilde{\Phi}(0)$ can be defined as $\lim_{\omega \to 0} \tilde{\Phi}(\omega)$ using continuity.  Secondly, the requirement $c_3 = 0$ is not invariant under special conformal transformations such as the inversion $\lambda \to 1/\lambda$.}

None of the zero mode constraints are physically important when proving the GSL.  That is because they relate to infrared issues on the horizon---to modes which are very long wavelength with respect to $\lambda$.  In other words, they relate to the behavior of the fields at $\lambda \to \pm \infty$.  But the GSL has to do with the relationship between two horizon slices at finite values of $\lambda$.  Any information which can only be measured at $\lambda = -\infty$ is totally irrelevant because it does not appear above either horizon slice.  On the other hand, information stored at $\lambda = +\infty$ can without loss of generality be equally well regarded as present in the asymptotic region $\mathcal{I}^+$ which `meets' the horizon at $\lambda = +\infty$.

Consequently the zero modes can simply be ignored.  This is a relief because zero mode issues tend to be one of the trickier aspects of quantum field theory on a null surface \cite{burkardt96}.  Since the mass $m$ only matters for calculating the zero mode and finite energy constraints, it will not be of significance for anything that follows.

\subsection{Smearing the Field}\label{smear}

Now $\Phi(x)$ is not a \textit{bona fide} operator, because the value of a field at a single point undergoes infinite fluctuations and therefore does not have well-defined eigenvalues (even though its expectation value $\langle \Phi(x) \rangle$ is well-defined for a dense set of states).  In order to get an operator, we need to smear the field in some $n$ of the $D$ dimensions with a smooth quasi-localized test function $f$:
\begin{equation}
\Phi(f) = \int f \Phi\,d^n x
\end{equation}
Because free fields are Gaussian, a finite width probability spectrum is sufficient to show that the operator is well-behaved.  So to check that $\Phi(f)$ has finite fluctuations, one can look to see whether its mean square $\langle \Phi(f)^2 \rangle$ is well-defined in the vacuum state.  Since spacetime is locally Minkowskian everywhere, the leading-order divergence can be calculated in momentum space using the Fourier transform of the smearing function $\tilde{f}$.
Because $f(x)$ is smooth, $\tilde{f}$ falls off faster than any polynomial at large $p$ values in all dimensions in which it is smeared, while it is constant in all the other dimensions.
  Up to error terms associated with $m^2$ and the curvature (whose degree of divergence must be less by 2 powers of the momentum), the fluctuations in $\Phi$ are thus given by:
\begin{equation}\label{pE}
\langle \Phi(f)^2 \rangle \propto \int d^D p\,\delta(p^2) H(p_0) \tilde{f}^2(p)
= \int_{E = |p|} \frac{d^{D-1}p}{2E}\,\tilde{f}^2(E, p),
\end{equation}
where $H$ is the Heaviside step function.  This means that in order to damp out the divergences coming from large $p$ values, it is sufficient to smear either in all the space directions or in the time dimension.  But neither of these is convenient for a null quantization procedure.  Instead one wants to be able to smear the integral in a null plane.  To do this we rewrite Eq. (\ref{pE}) in a null coordinate system $(p_u, p_v, p_y)$ where $y$ represents all transverse directions.  The mass shell condition is
\begin{equation}\label{nullshell}
p_v = \frac{p_y^2 + m^2}{p_u},
\end{equation}
and the integral over the lightcone (again neglecting mass and curvature) is
\begin{equation}\label{spectral}
\langle \Phi(f)^2 \rangle \propto
\int_{p_u p_v = p_y^2} d^{D-2}p_y\,H(p_u) \frac{dp_u}{p_u} \tilde{f}^2(p_v, p_y),
\end{equation}
where $f$ is smeared in the $v$ and $y$ dimensions but not in the $u$ dimension.  The integral is dominated by momenta that point nearly in the $p_u$ direction.  It falls off like $1/p_u$ for large $p_u$, so it is logarithmically divergent.  Therefore $\Phi$ does \emph{not} make sense as an operator when restricted to a horizon.

However, $\nabla_k \Phi$ does make sense as an operator, since its mean square has two extra powers of the null energy $p_v$ (one for each derivative):
\begin{equation}
\langle [\nabla_k \Phi(f)]^2 \rangle \propto
\int_{p_u p_v = p_y^2} d^{D-2}p_y\,H(p_u)\frac{dp_u}{p_u} p_v^2 \tilde{f}^2(p_v, p_y).
\end{equation}
By substituting in Eq. (\ref{nullshell}), this integral becomes
\begin{equation}
\int_{p_u p_v = p_y^2} d^{D-2}p_y\,H(p_u)\frac{dp_u\,p_y^4}{p_u^3} \tilde{f}^2(p_u, p_y)
\end{equation}
which is convergent.  (This may seem surprising, because taking derivatives normally makes fields more divergent, not less.  The extra factors of $p_v$ do make the integral more divergent in the $v$ direction, but that direction is already very convergent because of the rapid falloff of $\tilde{f}$.)

Since $\nabla_k \Phi(f)$ is a genuine operator, it generates an algebra $\mathcal{A}(H)$ on the horizon.

\subsection{Determinism}

Specifying $\Phi$ on $H$ is \emph{almost} enough to determine the value of $\Phi$ outside the horizon as well, by using Eq. (\ref{invert}) as a time evolution equation in the $u$ direction.  Since Eq. (\ref{invert}) is first-order in $\nabla_u$ it is not necessary to specify the velocities of the field, only their positions.  The reason it does not quite work is that $\nabla_v^{-1}$ is a nonlocal operator, making other boundary conditions potentially relevant.

Whether or not $\Phi$ can actually be determined is therefore a global issue depending on the causal structure of the whole spacetime.  In the case of a de Sitter horizon, $\Phi$ is determined by the value on $H$ since it is almost a complete Cauchy surface once one adds a single point at future infinity (the value of a free field should die away at late times, so the addition of this point doesn't change anything).  In the case of a Rindler horizon in Minkowski space the field is generically determined, since the only modes which are not determined are massless modes propagating in the exact same direction as the horizon.  But for a black hole horizon, the field $\Phi$ is not determined, since fields can also leave to future timelike or null infinity ($\mathcal{I}^+$).

Let $\Sigma$ be a complete Cauchy surface of the exterior of $H$, which includes both $H$ itself, and the asymptotic future $\mathcal{I}^+$ outside of $H$.  $H$ and $\mathcal{I}^+$ can be connected only at $\lambda = +\infty$.  However, any zero mode information measurable at $\lambda = +\infty$ can be assigned to the system $\mathcal{I}^+$.  In order to remove this redundant information from $H$, one can write the field at one time as the boundary term in an integral:
\begin{equation}\label{break}
\Phi(\lambda) =  \Phi(+\infty) - \int^{+\infty}_{\lambda} \nabla_k \Phi\,d\lambda^\prime,
\end{equation}
showing that classically, all the information in $\Phi(\lambda)$ not measurable at $\lambda = +\infty$ is stored in the derivative $\nabla_k \Phi$.  And this derivative, as shown in section \ref{smear}, is a well defined operator after smearing with a test function.

Thus the algebra of the whole spacetime can therefore be factorized into $\mathcal{A}(H) \otimes \mathcal{A}{(\mathcal{I}^+)}$, ignoring any degrees of freedom in the zero modes.

This means that there also exist states that factorize:
\begin{equation}\label{factor}
\Psi(\Sigma) = \Psi[\Phi(H)] \otimes \Psi[\Phi(\mathcal{I}^+)]
\end{equation}
The existence of these factor states is needed for the validity of the proof of the GSL in section \ref{outside}.  If there are any operators in the algebra which depend on the zero modes of $\Phi$, these may be considered part of the algebra of $\mathcal{I}^+$.

\subsection{Commutation Relations}

Ordinarily we are used to quantizing a scalar field with equal-time canonical commutation relation:
\begin{equation}
[\Phi(x_1),\,\dot{\Phi}(x_2)] = i\hbar \delta^{D-1}(x_1 - x_2).
\end{equation}
On a curved spacetime this relation can be covariantly adapted to any spacelike slice $\Sigma$ by using the determinant of the spatial metric $q$ and $\Sigma$'s future orthonormal vector $n^a$:
\begin{equation}\label{spacecom}
[\Phi(x_1),\,\nabla_n \Phi(x_2)] = i\hbar \delta^{D-1}(x_1 - x_2)/ \sqrt{q}.
\end{equation}
In order to obtain the commutation relations on a null surface, one can take the limit of an infinitely boosted spacelike surface.  Measured in any fixed coordinate system, each side of Eq. (\ref{spacecom}) diverges like $1/\sqrt{1 - v^2}$ due to the Lorentz transformation of $n^a$ or $1/\sqrt{q}$.  By dividing out the common divergent factor as one takes the limit, one ends up with\footnote{A previous version failed to include the factor of (1/2) in the following commutator, but a more careful derivation of the infinite boost limit shows that it is present.  This is easiest to see for massless fields in 1+1 dimensions, where the (1/2) appears because only left-moving modes contribute.}
\begin{equation}\label{nullcom}
[\Phi(y_1,\,\lambda_1),\,\nabla_k \Phi(y_2,\,\lambda_2)] =
\frac{i\hbar}{2} \delta^{D-2}(y_1 - y_2) \delta(\lambda_1 - \lambda_2)/ \sqrt{h}
\end{equation}
where $h$ is the determinant of the $D - 2$ spatial components of the horizon metric.  From now on, the factors of $1/\sqrt{g}$ or $1/\sqrt{h}$ will be automatically be absorbed into the definition of the delta functions $\delta^{D-1}(x)$ or $\delta^{D-2}(y)$ respectively.

By integrating Eq. (\ref{nullcom}) in the $\lambda_1$ direction, one can find the commutator of $\Phi$ with itself in terms of the Heaviside step function $H$:
\begin{equation}\label{phicom}
[\Phi(y_1,\,\lambda_1),\,\Phi(y_2,\,\lambda_2)] =
\frac{i\hbar}{2} \delta^{D-2}(y_1 - y_2)
[H(\lambda_2 - \lambda_1) - H(\lambda_1 - \lambda_2)]/2,
\end{equation}
where because the constant of integration only affects the zero modes, I have chosen it so that the commutator is antisymmetric.\footnote{One should not attempt to use Eq. (\ref{phicom}) in situations where zero modes are important, because then the constant of integration is undefined. This happens because the commutator of the full spacetime theory is ill-defined for null separations.  The reason Eq. (\ref{phicom}) can be used for the horizon theory is because all horizon observables will ultimately be expressed in terms of $\nabla_k \Phi$.}

Notice how even though the null surface acts like an initial data slice, there are nontrivial commutation relations of $\Phi$ on the horizon.  Since neither the commutation relations nor the generator of local null translations $T_{kk}$ carry any derivatives in the space directions, the horizon theory satisfies Ultralocality---i.e. the horizon theory is just the integral over a bunch of independent degrees of freedom for each horizon generator.

\subsection{Fock Space Quantization}\label{fock}

In order to perform Fock quantization, the fields will be analyzed in terms of the modes $\tilde{\Phi}$ with definite null-frequency $\omega$:
\begin{equation}
\tilde{\Phi}(y,\,\omega) = \int
\frac{e^{-i\omega \lambda}}{\sqrt{2\pi}} \Phi(y,\,\lambda)\,d\lambda,
\end{equation}
taking $\omega \ne 0$ in order to ignore the zero modes.  Because of Ultralocality, it is possible to define a Fock representation even when $y$ is kept in the position basis.

The commutation relations of the field in this basis can be calculated by taking the Fourier transform of Eq. (\ref{phicom}):
\begin{equation}
[\tilde{\Phi}(y_1,\,\omega_1),\,\tilde{\Phi}(y_2,\,\omega_2)] =
\hbar \frac{\delta(\omega_1 + \omega_2)}{\omega_2 - \omega_1} \delta^{D-2}y
\end{equation}
One can use this to define creation and annihilation operator densities
\begin{equation}
a^{\dagger}(y,\,\omega) = \tilde{\Phi}(y,\,\omega) \sqrt{\frac{2\omega}{\hbar}}, \quad
a(y,\,\omega) = \tilde{\Phi}(y,\,-\omega) \sqrt{\frac{2\omega}{\hbar}},
\end{equation}
which create and destroy particles of any frequency $\omega > 0$, and satisfy the commutation relations
\begin{equation}
[a(y_1,\,\omega_1),\,a^\dagger(y_2,\,\omega_2)] =
\delta(\omega_1 - \omega_2) \delta^{D-2}(y_1 - y_2).
\end{equation}
The single particle Hilbert Space corresponds to normalizable wavefunctions in the space $\Psi(y,\,\omega)$ ($\omega > 0$) of creation operators.  By taking the Fock space, one constructs the full Hilbert space of the scalar field on the horizon.

Because $T_{kk}$ is quadratic in the free field $\Phi$, the divergent part of the null energy $p_k$ is a state-independent constant.  In order to be Lorentz invariant the Hartle-Hawking vacuum $|0\rangle$ must have $p_k = 0$, so any physically reasonable renormalization of
$p_k$ (e.g. point-splitting) is equivalent to simply subtracting off the zero-point energy of the vacuum state.  Hence the renormalized null energy of the state can be calculated by rewriting Eq. (\ref{nullE}) in terms of the normal-ordered creation and annihilation operators:
\begin{equation}
p_k =
\int^{\infty}_{\omega=-\infty} \!\!\!\!\!\! \omega^2 \!:\!\tilde{\Phi}^*\tilde{\Phi}\!: d\omega\,d^{D-2}y
= \int^{\infty}_{\omega=0} \hbar\omega\,a^{\dagger}a\,d\omega\,d^{D-2}y
= \sum_n \hbar \omega_n,
\end{equation}
where the last equality is evaluated in the Fock basis of states which have a definite number of quanta of frequency $\omega_1 \ldots \omega_n$.  Thus the particles satisfy the Planck quantization formula.

The resulting picture of the scalar field theory on the horizon is surprisingly simple: each state is simply a superposition of a finite number of particles localized at distinct positions on the horizon, each with some positive amount of null energy $\hbar \omega$.  In contrast to the usual quantization on a spacelike surface, each particle can be arbitrarily well-localized near any horizon generator.  The particles cannot however be localized with respect to the $\lambda$ coordinate on the horizon generator.  No two particles can reside on exactly the same horizon generator, because that would not be a normalizable vector in the Fock space.

There is an enormous amount of symmetry of the scalar field theory on the horizon.  The only geometrical structures used in the quantization are the affine parameters of each horizon generator (up to rescaling), and the area-element (coming in via the $d^{D-2}y$ integration), which comes in through the commutation relation (\ref{nullcom}).  Therefore the Fock space is invariant under 1) arbitrary translations and dilations of the affine parameter of each horizon generator independently, 2) area-preserving diffeomorphisms acting on the space of horizon generators, and even 3) any non-area-preserving diffeomorphism that sends $d^{D-2}y \to \Omega(y)^2 d^{D-2}y$ so long as one also sends $\Phi \to \Omega(y)^{-1} \Phi$.  This is so much symmetry that the only invariant quantity is the total number $n$ of particles; every n-particle subspace of the Hilbert space is a single irreducible representation of the group of symmetries.\footnote{To see that this is the case, note that every $n$-particle state can be written as a superposition of states in which each of the $n$ identical particles is localized in a delta function on $n$ different horizon generators.  All such states are equivalent to one another by the symmetry transformations, so pick one of them, $\Psi$.  If the $n$-particle representation were reducible, there would have to exist a projection operator which is invariant under all the symmetry and acts nontrivially on this state by turning it into a linearly independent state $\Psi^\prime$.  But by virtue of the symmetry, $\Psi^\prime$ must be zero except on the $n$ horizon generators initially chosen, and therefore linearly dependent on $\Psi$.  Consequently the projection operator does not exist and the representation is irreducible.}

\subsection{Conformal Symmetry}\label{conf}

Even this does not exhaust the symmetries of the scalar field on the horizon (minus zero modes); one is actually free to perform any special conformal transformation of each $\lambda(y)$, i.e. any combination of a translation, dilation, and inversion $\lambda \to 1/\lambda$.  It is easiest to see this if the quantization is done in a slightly different way: by discretizing the horizon into a finite number of horizon generators.  Let there be $N$ discrete horizon generators spread evenly throughout the horizon area $A$, and let the field $\Phi(n,\,\lambda)$ be defined only on this discretized space.  The commutator is
\begin{equation}
[\Phi(m,\,\lambda_1),\,\nabla_k \Phi(n,\,\lambda_2)] =
\frac{i\hbar}{2}\frac{A}{N}\,\delta_{mn} \delta(\lambda_1 - \lambda_2),
\end{equation}
and the null energy is
\begin{equation}
p_k = \sum_{n = 1}^N \frac{A}{N} \int (\nabla_k \Phi_n)^2 \, d\lambda.
\end{equation}
These expressions converge to Eq. (\ref{nullcom}) and (\ref{nullE}) respectively as $N \to \infty$.  Since the theory is ultralocal there are no divergences associated with the transverse directions, so the limit should exist.  Every continuum horizon state can be described as the $N \to \infty$ limit of a sequence of states in the discretized model.  However, not every smooth seeming limit of states in the discretized model corresponds to a state in the continuum model; for example,
there is no continuum limit
of states in which one horizon generator has two particles on it and the rest are empty.

The discretized model is nothing other than a collection of $N$ different conformal field theories each of which is the left-moving sector of one massless scalar field in $1+1$ dimensions.  The entanglement entropy divergence is therefore just the same as in a conformal field theory (CFT) with $N$ scalar fields, which has central charge $c = N$ \cite{ginsparg89}:
\begin{equation}
S_\mathrm{div} = \frac{c}{12} \ln \left( \frac{\lambda_I}{\lambda_U} \right)
\end{equation}
where $\lambda_I$ is the affine distance of the infrared cutoff from the boundary, and $\lambda_U$ is the affine distance of the ultraviolet cutoff.  This justifies Eq. (\ref{nulldiv}) mentioned in section \ref{ren} on renormalization.

In any CFT, the vacuum state $| 0 \rangle$ is invariant under all special conformal transformations.  Since the $N \to \infty$ limit of $| 0 \rangle$ is just the vacuum of the continuum theory, the continuum vacuum is also invariant under the group of special conformal transformations $SO(2,\,1)$.

A $1+1$ dimensional CFT is also invariant under general conformal transformations, i.e. arbitrary reparameterizations of a null coordinate $v \to f(v)$.  However, the vacuum state is not invariant under general conformal transformations.  This is a consequence of the anomalous transformation law of the stress energy tensor $T_{vv}$ \cite{ginsparg89}:
\begin{equation}
T_{vv} \to f^\prime(v)^{-2} T_{vv} + \frac{c}{12} S(f),
\end{equation}
where $c = 1$ is the central charge of one scalar field, and $S(f)$ is the Schwarzian derivative:
\begin{equation}
S(f) =
\frac{f^{\prime\prime\prime}}{f^\prime} -
\frac{3}{2}\frac{(f^{\prime\prime})^2}{(f^\prime)^2},
\end{equation}
which vanishes only when $f(v)$ is special.  Since the vacuum must have $T_{vv} = 0$, any nonspecial conformal transformation of the vacuum must produce a nonvacuum state with positive expectation value of the null energy $p_k$.

What if one tries to perform a general conformal transformation $\lambda \to f(\lambda,\,y)$ of the horizon generator parameters $\lambda$ for $D > 2$ dimensions?  In the discretized model, the null energy of the transformed vacuum is
\begin{equation}
p_k = \sum_{n = 1}^N \frac{1}{12} \int S(f,\,n) d\lambda
\end{equation}
and the integrand is positive.  But now disaster strikes---as $N \to \infty$, $p_k \to \infty$ too!  The general conformal transformation takes the vacuum out of the Hilbert space altogether, by creating infinitely many quanta.  So the conformal anomaly prevents $\lambda$ from being reparameterized, except by a special conformal transformation.

Since the stress-energy $T_{kk}$ is the generator of reparameterizations, this means that most integrals of $T_{kk}$ on the horizon do not give rise to operators in the Hilbert space.  Since $T_{kk} = (\nabla_k \Phi)^2/2$ is a product of two fields, there is a danger of divergence.  The fact that only special conformal transformations of the vacuum are allowed implies that the only integrals of $T_{kk}$ which are horizon observables are those of this form:
\begin{equation}
\int^{+\infty}_{-\infty} T_{kk}\,[a(y) + b(y)\lambda + c(y)\lambda^2]\,d\lambda\,d^{D-2}y.
\end{equation}
For example, the restricted boost energy
\begin{equation}
K(\Lambda) = \int_\Lambda^\infty T_{kk}\,(\lambda - \Lambda) \,d\lambda\,d^{D-2}y
\end{equation}
is not an operator because of the limitation of the integral to $\lambda > \Lambda$.  However,
the proof is only concerned with the expectation value $\langle K(\Lambda) \rangle$.  This is a function of $\langle T_{kk}(x) \rangle$, which does not need to be smeared to be finite.

\section{Other Spins}\label{spin}

In this section some basic details of null quantization for alternative spins will be briefly provided, omitting detailed derivations and neglecting zero modes.

\subsection{Spinors}

The Lagrangian of any free spinor field can be written as
\begin{equation}\label{spinorL}
\mathcal{L} =  \gamma^{ABi} \Psi_A \nabla_i \Psi_B + m \epsilon^{AB} \Psi_A \Psi_B,
\end{equation}
where $A$ or $B$ belong to spinor representations written in a real (Majorana) basis, $\gamma^{ABi}$ is the gamma matrix, and $\epsilon^{AB}$ is the invariant symplectic structure on the spinor space.\footnote{In dimensions $D\,\mathrm{mod}\,8 = 0,\,1,\,2,\,6$, the irreducible spinor representations do not possess an invariant symplectic structure $\epsilon^{AB}$.  Consequently, for $m > 0$ it is necessary to use reducible spinor representations.  The Majorana spinor basis has been chosen in order to keep the spinor expressions homogeneous across different spacetime dimensions.  Dirac and/or Weyl spinors may be obtained from representations which admit a complex structure.} As long as $D > 2$, the qualitative features of null surface quantization are the same for every kind of spinor.\footnote{In $D = 2$, the chirality of the field determines whether it propagates to the left or to the right.  Only fields which propagate across a null surface can be quantized on that surface.}

The equation of motion is
\begin{equation}\label{maj}
\nabla_i \Psi_B \gamma^{ABi} = m \Psi^A,
\end{equation}
using $\epsilon^{AB}$ to raise the spinor index.  At any point on a spacelike slice of the horizon, the $D$ dimensional spinor decomposes into the tensor product of a Majorana spinor in $D - 2$ dimensional space, and a Dirac spinor on a $1 + 1$ dimensional spacetime.  The Dirac spinor in $1 + 1$ dimensions decomposes into the direct sum of a left-handed spinor $\Psi_L$ and a right-handed spinor $\Psi_R$, where we take $\gamma^{RRa}$ to point in the $k^a$ direction and $\gamma^{LLa}$ to point along the other lightray $l^a$.  The Majorana equation (\ref{maj}) takes the schematic form:
\begin{eqnarray}
\nabla_{LL} \Psi_R + \nabla_{LR} \Psi_L + m \Psi_L
= \nabla_k \Psi_R + \nabla_y \Psi_L + m \Psi_L \label{L}; \\
\nabla_{RR} \Psi_L + \nabla_{RL} \Psi_R + m \Psi_R
= \nabla_l \Psi_L + \nabla_y \Psi_R + m \Psi_R.\label{R}
\end{eqnarray}
The first equation (\ref{R}) only involves derivatives that lie on the horizon itself, and can be used to define $\Psi_R$ as a function of $\Psi_L$ (up to zero modes):
\begin{equation}
\Psi_R(\lambda) =
\Psi_R(+\infty) - \int_\lambda^{+\infty} (\nabla_y \Psi_L + m \Psi_L)\,d\lambda^\prime.
\end{equation}
On the other hand, Eq. (\ref{L}) determines the derivative of $\Psi_L$ off the horizon, and so it does not act as a constraint.  Therefore, the spinor degrees of freedom are determined by the arbitrary specification of $\Psi_L(y,\,\lambda)$ on the horizon.  From now on we will focus on just the $\Psi_L(y,\,\lambda)$ degrees of freedom.

$\Psi_L(y,\,\lambda)$ yields a (fermionic) operator when smeared over the horizon directions by a test function $f$.  The mean-square of a massless spinor in momentum space is
\begin{equation}
\langle \Psi_L(f)^2 \rangle \propto
\int_{p_u p_v = p_y^2} d^{D-2}p_y\,H(p_u)\frac{dp_u}{p_u} p_v \tilde{f}^2(p_u, p_y).
\end{equation}
The extra power of $p_{LL} = p_v = (p_y^2 + m^2)/ p_u$ comes from the contraction of the momentum with the spin in the propagator, and serves to render the integral convergent.  Thus for spinors there is no need to take a $\nabla_k$ derivative in order to restrict the field to the horizon.

The anticommutator of the field on a spatial slice $\Sigma$ with normal vector $n^a$ is:
\begin{equation}
\{ \Psi_A(x_1) ,\, \Psi_B(x_2) \} = -i\hbar\,\gamma_{ABn} \delta^{D-1}(x_1 - x_2).
\end{equation}
By making an infinite boost, one can obtain the anticommutator for the field $\Psi_L$ on the horizon:
\begin{equation}
\{ \Psi_{IL}(y_1,\,\lambda_1) ,\, \Psi_{JL}(y_2,\,\lambda_2) \} =
-\frac{i\hbar}{2} g_{IJ} \delta(\lambda_1 - \lambda_2) \delta^{D-2}(y_1 - y_2),
\end{equation}
where $I$ and $J$ are (real) spinor representations of $SO(D - 2)$ (the group of rotations of the $D - 2$ dimensional transverse space).  Since these representations are unitary, there is a natural metric $g^{IJ} = \gamma^{ILJL}_k$ on the transverse spinor space.

The null-null component of the stress-energy is\footnote{The stress-tensor is easiest to calculate canonically by contracting the null momentum (\ref{conpi}) by the velocity $\nabla_k \Psi_L$.  The gravitational $T_{kk}$ is the same, but calculating it requires intorducing an $n$-bein, and varying the Lagrangian (\ref{spinorL}) with respect to it.}
\begin{equation}
T_{kk} = -g^{IJ} \Psi_{IL} \nabla_k \Psi_{JL}.
\end{equation}
$T_{kk}$ and the anticommutation relations look just like the integral of the corresponding quantities for left-moving spinor fields in $1+1$ dimensions.  Therefore, if the horizon generators are discretized, the corresponding CFT is that of $N/2$ massless left-moving chiral fermions, where $N$ is the number of components of the spinor field.

\subsection{Photons}\label{phot}

The Maxwell Lagrangian is $\mathcal{L} = -\frac{1}{4}F_{ab}F^{ab}$.  It is convenient to impose Lorenz gauge $\nabla_a A^a = 0$ and null gauge $A_k = 0$.  Let $i$ be the transverse directions restricted to the horizon, and let $l$ be a null direction pointing away from the horizon, such that $g_{kl} = -1$ and $g_{il} = 0$.  By integrating the Lorenz gauge, one can solve for $A_l$ (up to zero modes) in terms of the transverse components $A_i$:
\begin{equation}
A_l = A_l(+\infty) -\int^{+\infty}_\lambda \! \nabla_i A^i \,d\lambda,
\end{equation}
where we have used null gauge and the fact that $R_{klki}$ vanishes on a stationary horizon.  Hence the only independent (nonzero-mode) degrees of freedom are the transverse directions $A_i$ on the horizon.  The commutator is
\begin{equation}
[A_i(y_1,\,\lambda_1), \nabla_k A_j(y_2,\,\lambda_2)] =
\frac{i\hbar}{2} g_{ij} \delta^{D-2}(y_1 - y_2) \delta(\lambda_1 - \lambda_2),
\end{equation}
and the stress-energy tensor is
\begin{equation}
T_{kk} = g^{ij} (\nabla_k A_i) \nabla_k A_j.
\end{equation}
$A_i$ cannot be smeared to make a valid operator on the horizon, but $\nabla_k A_i$ can.

After discretization of horizon generators, the CFT of each horizon generator consists of $D - 2$ left-moving massless scalars.

\subsection{Gravitons}\label{grav}

In the semiclassical limit the metric can be described as a background metric
$g_{ab} \equiv g_{ab}^0$ plus an order $\hbar^{1/2}$ metric perturbation $h_{ab} = g_{ab}^{1/2}$.  Impose Lorenz gauge $\nabla_a h^a_b - \frac{1}{2} \nabla_a h^b_b = 0$ and null gauge $h_{ka} = 0$.

The Lagrangian and equations of motion are simply that of perturbative general relativity (GR).  The only constraint on $h_{ab}$ on the horizon at half order is the null-null component of the Einstein equation:
\begin{equation}\label{Gkk}
G_{kk} = 0.
\end{equation}
By integrating $\nabla_k \theta^{1/2} = 0$ (the half order Raychaudhuri equation (\ref{Rayhalf}), one finds that there is no half order contribution to the area:
\begin{equation}\label{freeze}
h_{ij} g^{ij} = 0.
\end{equation}
In order to keep things simple, the trace degree of freedom of $h_{ij}$ will therefore be set to zero before quantization.  Only the traceless part of $h_{ij}$ represents physical graviton degrees of freedom.\footnote{Rotational symmetry assures that the commutator of the trace degrees of freedom cannot mix with the commutator of the traceless degrees of freedom.  The constraint (\ref{Gkk}) generates diffeomorphisms in the $k$ direction.  Consequently if one wished to impose this constraint after quantization, for consistency it would also be necessary to include as a physical degree of freedom the parameter $\lambda$ which breaks this symmetry.}

$h_{ij}$ cannot be smeared to make an operator on the horizon, but $\nabla_k h_{ij}$ can.  Thus, the only physical components of the field are the transverse shear components $\sigma_{ij} \propto \nabla_k h_{ij}$.

In GR, gravitons do not contribute to the gravitational stress-energy tensor $T_{ab}$ found by varying the matter Lagrangian with respect to the metric, since gravitons do not contribute to the matter Lagrangian.  And if one varies with respect to the full gravitational Lagrangian, the resulting tensor vanishes when the equations of motion are satisfied.  However, in perturbative GR, one can still define a stress-energy tensor perturbatively by varying the Lagrangian with respect to the \emph{background} metric, rather than the perturbed metric.  The resulting stress-energy tensor is proportional to the contribution of $h_{ab}$ to the Einstein tensor:
\begin{equation}
T_{ab}^1 = G_{ab}^{1} / 8\pi G,
\end{equation}
to first order in $\hbar$.  On the horizon this is just
\begin{equation}\label{gravT}
T_{kk} = (\nabla_k h_{ij}) \nabla_k h^{ij} / 32\pi G.
\end{equation}

The canonically conjugate quantities for canonical general relativity on a spacelike slice $\Sigma$ are the spatial metric $q_{ab}$ and the extrinsic curvature
$K_{ab} = \nabla_{n} q_{ab}/2$ \cite{ADM}:
\begin{equation}
[q_{ab}(x_1),\,(K^{cd} - q^{cd} K)(x_2)] = \frac{i\hbar}{2} (8\pi G) \frac{\delta_a^c\delta_b^d + \delta_b^c\delta_a^d}{2} \delta^{D-1}(x_1 - x_2)
\end{equation}
If one takes the infinite boost limit, the spatial extrinsic curvature $K_{ij}$ with $i,\,j$ lying in the transverse plane becomes the null extrinsic curvature:
\begin{equation}
K_{ij} \to B_{ij} = \nabla_k h_{ij}/2 = \sigma_{ij} + \frac{1}{D-2} g_{ij} \theta.
\end{equation}
Because the trace part has been made to vanish by Eq. (\ref{freeze}), only the traceless shear part remains.  Therefore the commutator is
\begin{equation}
[h_{ij}(y_1,\,\lambda_1),\, \sigma^{lm}(y_2,\,\lambda_2)] = \frac{i\hbar}{2} (8 \pi G)
\delta_{ij}^{lm} \delta^{D-2}(y_1 - y_2) \delta(\lambda_1 - \lambda_2),
\end{equation}
where $\delta^{lm}_{ij} = \frac{1}{2}(\delta^l_i \delta^m_j + \delta^l_j \delta^m_i)
- \frac{1}{D-2} g_{ij}g^{lm}$ is the Kroneker delta for the traceless symmetric representation.

As for the other bosonic fields, $\sigma_{ij}$ is an observable when smeared on the horizon, but $h_{ij}$ is not.  When the horizon generators are discretized, the graviton CFT is that of $\frac{1}{2}(D^2 - 3D)$ left-moving scalar fields.

\section{Interactions}\label{int}

Does the argument given in section \ref{arg} for the GSL continue to work when the quantum fields have nontrivial interactions besides the minimal coupling to gravity?  The question is whether one can continue to define a horizon algebra $\mathcal{A}(H)$ satisfying the four axioms required for the proof described in sections \ref{outline} and \ref{sym}: Determinism, Ultralocality, Local Lorentz Invariance, and Stability.  Except for free fields and 1+1 CFT's (see below), it is not obvious that this is the case.  Some evidence for and against the existence of such an algebra will be presented below.  Hopefully future work will clarify these issues.

\subsection{Perturbative Yang-Mills and Potential Interactions}

Let $\phi_i$ stand for a field (indexed by $i$) in any free field theory, of any spin.  What happens to the horizon algebra upon adding interactions?

In general, the addition of arbitrary terms to the Lagrangian will change both the commutation relations and the value of the null stress-energy tensor $T_{kk}$.  But for certain special kinds of interactions, the null algebra may remain unaffected.

In particular, at least at the level of formal perturbation theory, the horizon fields $\phi_i$ do not care about the addition of an arbitrary potential term $V(\phi)$ to the Lagrangian.  In order to be a potential, $V$ must depend only on the fields and the metric, not field derivatives or the Riemann tensor.

The general horizon commutator can be written as
\begin{equation}
[\phi_i ,\, \Pi^i] = \frac{i\hbar}{2} \delta^{D-2}(y_1 - y_2) \delta(\lambda_1 - \lambda_2),
\end{equation}
where the conjugate momentum to the field in the null direction is given by
\begin{equation}\label{conpi}
\Pi^i = -\frac{ \partial \mathcal{L} }{\partial \nabla^a \phi_i } k^a,
\end{equation}
and the commutator is replaced with an anticommutator for fermionic fields.  $V$ does not depend on any derivatives of the field:
\begin{equation}
\frac{ \partial V}{\partial \nabla^a \phi_i } = 0,
\end{equation}
so the momentum $\Pi^i$ is the same as in the free theory.  Since the horizon algebra is generated by the free field operators subject to the above commutation relation, the horizon algebra $\mathcal{A}(H)$ is unaffected by the perturbation.

A similar result holds for Yang-Mills interactions.  The Yang-Mills Lagrangian coupled to spinors and scalars is
\begin{equation}
\mathcal{L} = -\frac{1}{4} F_{ab} F^{ab} - \frac{1}{2} \nabla_a \Phi \nabla^a \Phi + \gamma^{ABi} \Psi_A \nabla_i \Psi_B,
\end{equation}
where $F_{ab} = \nabla_a A_b - \nabla_b A_a$.  Because $\nabla_a$ is the covariant derivative, there are cubic boson interactions which depend on the $\nabla^k$ derivative, of the form $A^a A_k \nabla^k A_a$ and $A_k \Phi \nabla^k \Phi$.  However, these interactions both depend on $A_k$, which vanishes in null gauge (which was used to obtain the horizon algebra in section \ref{phot}).  The spinor interactions do not depend on $\nabla_k$.  So Yang-Mills interactions also do not affect $\mathcal{A}(H)$, as a special consequence of gauge symmetry.

Because the horizon algebra is the same, the generator of null translations $T_{kk}$ must also be the same.  Since for minimally coupled theories the canonical stress-tensor and the gravitational stress-tensor of matter are the same up to boundary terms at infinity \cite{fursaev99}, this means that the formula for the area $A$ in terms of $T_{kk}$ is the same.  Also, the (translation-invariant) vacuum state $|0\rangle$ of the interacting field theory is the same as the free field vacuum, up to zero modes \cite{burkardt96}.  This is because, unlike spatial surfaces, null surfaces have a kinematic momentum operator $p_k$ which is required to be positive.\footnote{In the case of spacelike surfaces, the interacting vacuum cannot even lie in the Fock space of the free vacuum \cite{EF05}.}  Since everything in $\mathcal{A}(H)$ is exactly the same as in the free case, at the level of formal perturbation theory the entire proof goes through without depending in any way on the interactions.

However, this entire discussion needs to be taken with a large grain of salt, because it assumes that the interactions in the Lagrangian can be treated as a finite perturbation.  Once loop corrections are taken into account, there will be divergences which have to be absorbed into the coupling constants.  Even if one starts with an interaction potential $V(\phi)$ which seems not to have any harmful derivative couplings in it, renormalization will typically produce derivative couplings which will affect the commutation relations.

For example, a field strength renormalization of the propagator term will change the overall coefficient of the commutation relation.  This field strength renormalization will usually be infinite---except when the theory is superrenormalizable.  So for e.g. Yang-Mills or $\Phi^4$ in $D = 3$, the above arguments suggest that the horizon algebra should be unaffected.  This however, has not been shown rigorously at the nonperturbative level.  For marginally renormalizable or nonrenormalizable theories, the horizon algebra might be deformed, or it might not exist at all.  Nevertheless, null quantization methods have been useful for ($D=4$) QCD calculations \cite{burkardt96},  notwithstanding the fact that it may not be rigorously justified.

In the case of spacelike hypersurfaces, there is a series of theorems \cite{powers67} which show that for any quantum field theory which is reducible to bosons and fermions satisfying the equal time canonical (anti-)commutation relations (ETCCR), the theory must be free unless the interactions are sufficiently weak in the ultraviolet.  Superrenormalizable theories do obey the ETCCR, nonrenormalizable theories cannot obey the ETCCR (even if they can be defined using a UV fixed point), while the status of marginally renormalizable theories is unclear.  The problem arises because of infinite renormalization of the fields.  Thus there exist at least some QFT's which do not satisfy the equal time ETCCR.  One \emph{possible} interpretation of this result is that the ``equal time'' is at fault, and it is necessary to smear the fields in time as well as in space in order to get a well defined operator.  This probably would mean that such fields are not well defined when smeared on a null surface either.  However, it could still be that there exist a different set of fields which do not obey canonical commutation relations, and can be defined on the horizon algebra.

\subsection{Conformal Field Theories}\label{nonpert}

So do nonperturbatively interacting QFT's really have a horizon algebra?  One can get some insight by studying conformal field theory (CFT).  Any physically consistent QFT must have good ultraviolet behavior as length scales are taken to zero.  The conventional wisdom is that this happens if and only if the theory approaches an ultraviolet fixed point of the renormalization group flow.  At short distances, the theory is therefore scale invariant.  All known scale invariant QFT's are also conformally invariant, so let us ask whether CFT's have a null surface formalism.  Since the near-horizon limit is a type of ultraviolet limit, it seems probable that a QFT has a null surface formulation if and only if the scaling limit CFT does.

The situation is very different for 1+1 CFT's (which have an infinite conformal group) and higher dimensional CFT's (which have a finite conformal group).

\paragraph{\textbf{1+1 CFT}.} In the case of 1+1 CFT's, there always exists a nontrivial algebra of observables $\mathcal{A}(H)$ on the horizon (i.e. on a lightray), which is simply the algebra of the left-moving chiral currents.  To see this, we remind the reader of some facts about 1+1 CFT's (from e.g. \cite{ginsparg89}).  The operators of a CFT fall into infinite dimensional representations of the conformal algebra associated with the theory's central charges $c$ and $\tilde{c}$.  These representations are classified by the weight spectrum of primary operators $(h,\,\tilde{h})$, which specify the weight of the primary operator in the representation with respect to left and right dilations.  Descendants of these operators have weights given by the primary operators plus integers.

The algebra of operators which are well-defined on the horizon is simply the algebra of left-moving chiral operators (i.e. the algebra generated by quasi-primary operators weight $(h,\,0)$).  Such fields do not depend on the $u$ coordinate and therefore must be localizable to the horizon.  (On the other hand, the two two-point function of a non-chiral operator diverges when the two points are null separated on the horizon, so such operators cannot be smeared in one null direction alone.)  Since the identity operator has weight $(0,\,0)$, there is always an infinite sequence of such operators, including the null stress-energy $T_{kk}$ of weight $(2,\,0)$.  Thus there is always an infinite nontrivial horizon algebra $\mathcal{A}(H)$, which includes the generators of the conformal group itself.

We now examine whether this horizon algebra obeys the necessary axioms described in section \ref{sym} for the proof of the GSL.  Ultralocality is trivial in 1+1 dimensions, since there is only one horizon generator.  Lorentz Symmetry and Stability hold by virtue of the normal QFT axioms.\footnote{Although the discussion in this subsection is entirely about QFT on a fixed background spacetime, the reader may wonder why one would want to consider a 1+1 CFT's for a matter sector given that GR is topological in 2 dimensions.  The answer is that the proof given in section \ref{arg} is equally applicable to 2d dilaton gravity, in which the dilaton plays the role of the  ``area''.}

The only tricky point is Determinism, which requires the exterior of the horizon to be determined by $\mathcal{A}(H)$ and $\mathcal{A}(\mathcal{I}^+)$.  In the case of a chiral CFT, which breaks into independent left-moving and right-moving sectors, Determinism is obvious.  But in the case of a non-chiral theory, there exist operators which are not products of chiral currents, e.g. the operators built on the weight $(1/16,\,1/16)$ and weight $(1/2,\,1/2)$ operators in the Ising model.  It seems possible that there might be information stored in the other operators, which cannot be measured strictly on the lightfront.

It seems however that information outside the horizon would still be contained in the algebra generated by (a) left-chiral operators in $\mathcal{A}(H)$, (b) right-chiral operators in $\mathcal{A}(\mathcal{I}^+)$, and (c) all operators (including non-chiral operators) in a neighborhood $N[i^+]$ of timelike infinity $i^+$, where $H$ joins onto $\mathcal{I}^+$.\footnote{By ``generated'', we mean that we consider the smallest von Neumann algebra containing all the operators in (a)-(c).}  That is because any operator in $\mathcal{A}(N[i^+])$ could be translated to any other point outside the horizon by use of the stress tensor current.  Thus, if we interpret Determinism in this way, then this appears to imply the GSL for general 1+1 dimensional CFT's.\footnote{This argument replaces an erroneous one that was given in previous versions of this article.}



\paragraph{\textbf{Higher dimensional CFT.}}  In higher dimensional interacting CFT's, a local field will no longer obey the free wave equation.  This means that it must have a nonzero anomalous dimension $\eta$.  For example, a primary scalar field in $D$ dimensions will have a dimension $\Delta = (D - 2)/2 + \eta$, with $\eta > 0$ due to the unitarity bound.  Such fields do not form operators when smeared on the horizon alone.  This can be seen from evaluating the square of the smeared field using the spectral decomposition of the operator:
\begin{equation}
\langle \Phi(f)^2 \rangle \propto
\int_{p^2 < 0} d^{D}p\,H(p_0) \frac{\tilde{f}^2(p_v, p_y)}{(-p^2)^{1 - \eta}},
\end{equation}
where $\tilde{f}$ is the Fourier transform of the smearing integral on the horizon.  This expression is the analogue of Eq. (\ref{spectral})), but now the integral is performed over all timelike momenta $p^2 < 0$.  Because of the smearing, the integral is dominated by momenta which point nearly in the $p_u$ direction.  Since $p^2 = p_y^2 - p_u p_v$, the integral falls off in the $p_u$ direction like $p_u^{\eta - 1}$.  This is divergent for all permitted values of $\eta$.  Consequently no operator can be defined.  Unlike the free case, it is no longer possible to improve the situation by taking $\nabla_v$ derivatives, since the $p_u$ and $p_v$ directions are no longer related by the null mass shell condition.

Similar arguments rule out operators formed from interacting fields with spin $\phi_I$, where $I$ transforms in a spin-$s$ irrep.  Let the conjugate field be written $\phi^*_{I^\prime}$.  In this case it is necessary (but not always sufficient) to satisfy the unitary bound that the primary have weight $\Delta = (D - 2)/2 + s + \eta$ for an $\eta > 0$ \cite{mack}.  The absolute square of the field smeared on the horizon looks like:
\begin{equation}
\langle \phi(f)_I \phi^* (f)_I^\prime \rangle \propto
\int_{p^2 < 0} d^{D}p\,H(p_0) \epsilon_{II^\prime}(p)
\frac{\tilde{f}^2(p_v, p_y)}{(-p^2)^{1 - s - \eta}},
\end{equation}
where $\epsilon_{II^\prime}(p)$ is the scalar product of the spins $I$ and $I^\prime$ in the little group $SO(D - 1)$ that preserves the momentum $p$.  At fixed $p_v$ and large $p_u$, $\epsilon$ can scale like $p_u^{2x}$ where $-s \le x \le s$ depends on the weight of the particular polarization under Lorentz boosts.  This integral is still divergent.  So it is also impossible to construct $\mathcal{A}(H)$ from fields of higher spin.

Nevertheless, this does not entirely rule out the possibility that there might be a nontrivial horizon algebra $\mathcal{A}(H)$, so long as it is constructed from operators that do not come from smearing local fields.  As an analogy, there exist CFT's in which fields cannot be defined by smearing on a $D - 1$ dimensional spacelike surface $\Sigma$.\footnote{This can be seen by doing a spectral decomposition of a primary scalar field with $\eta \ge 1/2$.}  Nevertheless, one can still define a local algebra on an incomplete spatial surface $\Sigma$ by means of the Hodge duality $\mathcal{A}(\Sigma) = \mathcal{A}^\prime(\Sigma^\prime)$, i.e. by defining $\mathcal{A}(\Sigma)$ to include any operator which commutes with all observables that are spacelike separated from $\Sigma$.  It may be that some similar trick can be used to define the observables on a null surface.

A possible argument that $\mathcal{A}(H)$ should exist is that in a CFT there is no distinction between finite and infinite distances.  Consequently, one can apply a Weyl rescaling $g_{ab} \to \Omega^2(x) g_{ab}$ with the property that the affine distance to the horizon becomes infinite.  Because curvature has mass dimension $2$, this also should lead to the scaling away of any curvature effects.  The existence of an algebra on the horizon is now equivalent to the existence of final scattering observables for particles travelling into this new, nearly flat asymptotic region.  This converts the ultraviolet problem of null restriction to the infrared problem of final scattering states.

However, because a CFT has no mass gap, there are long range interactions, and the asymptotic states might not form a Fock space, due to the possibility of creating an infinite number of soft massless particles.  In order to apply the proof of the GSL in section \ref{arg}, one would need to show that despite the existence of these long range forces, the final scattering algebra can be decomposed into a part associated with $H$ and a part associated with $\mathcal{I}^+$:
\begin{equation}
\mathcal{A}(H\,\cup\,\mathcal{I}^+) = \mathcal{A}(H) \otimes \mathcal{A}(\mathcal{I}^+)),
\end{equation}
and also show that $\mathcal{A}(H)$ obeys the other three axioms: Ultralocality, Local Lorentz Invariance, and Stability.

If there are any QFT's in which the algebra $\mathcal{A}(H)$ does not exist, extending the proof would presumably require a more delicate near-horizon limit.  One would have to show that a small smearing of fields out from the horizon does not break the symmetry group of the horizon sufficiently to spoil the proof.

\subsection{Higher Curvature and Nonminimal Coupling}\label{nonmin}

Further generalization of the proof is necessary when the gravity theory goes beyond the Einstein theory, either because the matter fields are nonminimally coupled, or because there are higher curvature terms in the gravitational Lagrangian.  In general, the presence of such terms will not only change the metric field equations, but also lead to the addition of extra terms in the horizon entropy $S_\mathrm{H}$.  These corrections can be calculated for stationary black holes by means of the Wald Noether charge method \cite{WI94}; however, there are certain ambiguities which arise for the case of dynamically evolving horizons.  Except for some special cases like $f(R)$ gravity (which can be related by field redefinitions to scalar fields minimally coupled to general relativity \cite{2ndlaw}) it is unknown whether such theories even obey a classical second law, let alone a generalized one.  For example, it appears that the Wald entropy can decrease when black holes merge in Lovelock gravity \cite{lovelock}.

Although the present work is restricted to the Einstein theory, some insight into these problems might be gained by analyzing the structure of horizon observables in non-Einstein theories.  The reason why the GSL holds on black holes in general relativity is that $\mathcal{A}(H)$ is small enough to have lots of symmetry (Local Lorentz Invariance) and yet large enough to contain all the information falling across the horizon (Determinism).  In general, alternative gravities will require $\mathcal{A}(H)$ to depend on additional information besides the metric and affine parameter on the horizon, e.g. curvature components.

If this additional information breaks the ability to translate each horizon generator independently, this may account for the failure of the second law in these theories.  Another reason why theories may fail to obey the second law is if the theory permits negative energy excitations, violating the Stability axiom.

On the other hand, if a horizon field theory for matter and gravitons can be found which still obeys all four axioms used in section \ref{arg}, this is auspicious for the GSL.  It might be that the ambiguities in the Wald Noether charge can be fixed by requiring that $S_\mathrm{H}$ depend only on quantities measurable in $\mathcal{A}(H)$ itself.  Suppose that this were done.  Then the GSL might be shown by the following argument:

First we need an analogue of Eq. (\ref{Tint}), relating the horizon entropy to the boost energy falling across the horizon:
\begin{equation}\label{Sint}
S_\mathrm{H}(\Lambda) = S_\mathrm{H}(+\infty)
- \frac{2\pi}{\hbar} \int_\Lambda^\infty \langle T_{kk} \rangle \,(\lambda - \Lambda) \,d\lambda\,d^{D-2}y.
\end{equation}
But the Wald Noether charge method shows that this is true in any classical diffeomorphism invariant theory when $T_{kk}$ is interpreted as a \emph{canonical} stress-energy current \cite{WI94}.  (The ``gravitational'' stress energy tensor defined by varying with respect to the metric is not very meaningful at this level of generality, because it is not invariant under field redefinitions of the metric).  Wald's argument is classical, so in order to use Eq. (\ref{Sint}), one would have to show that it survives a semiclassical quantization of the matter fields.

Since the canonical stress-energy tensor generates diffeomorphisms, one can also rewrite Eq. (\ref{Sint}) in terms of $K(\Lambda)$, the generator of boost symmetries about a horizon slice with $\lambda = \Lambda$:
\begin{equation}
S_\mathrm{H}(\Lambda) = C - 8\pi G\,\langle K(\Lambda) \rangle.
\end{equation}

Since the canonical stress-energy tensor is the generator $K$ of boost symmetries, by the Bisongano-Wichmann theorem, the quantum fields should be in a thermal state with respect to $K$.  Assuming that a non-Einstein gravity theory satisfies each of the criteria described above, it too should obey a semiclassical GSL.

\textbf{Note added v6:} Since this article was originally published, significant progress has been made.  \cite{Bousso:2014uxa} confirmed that in higher-dimensional interacting CFT's, there are indeed no operators localized in compact regions on null surfaces.  Nevertheless, \cite{Casini:2017roe,Lashkari:2017rcl} showed that the wedge regions \emph{outside} of arbitrary horizon slices have modular Hamiltonians $K$ whose action on a null plane looks like Eq. (\ref{modular}).  Together with the results of this paper, this implies that the GSL holds for general interacting QFT's.  Also, \cite{Wall:2015raa} extended this proof of the GSL to the case where the matter fields are coupled to an arbitrary higher-curvature gravity theory, assuming that the background spacetime is a regular Killing horizon.

\small
\subsection*{Acknowledgments}
I am grateful for comments by Ted Jacobson, Rafael Sorkin, Rob Myers, Will Donnelly, Sudipta Sarkar, Yoh Tanimoto, Renaud Parentani and Ed Witten.  Supported in part by the National Science Foundation grants PHY-0601800 and PHY-0903572, the Maryland Center for Fundamental Physics, the Simons Foundation, the Perimeter Institute, and the Institute for Gravitation and the Cosmos at Penn State.

\end{document}